\documentclass{aa}
\usepackage{txfonts}
\usepackage{graphicx}
\usepackage{t1enc}
\usepackage{epsfig}
\usepackage{rotating}
\usepackage[authoryear]{natbib}
\bibpunct{(}{)}{;}{a}{}{,}

\newcommand{\Tex}   {T_\mathrm{ex}}
\newcommand{\Trot}  {T_\mathrm{rot}}
\newcommand{\Tmb}   {T_\mathrm{MB}}

\newcommand{\kms}   {km~s$^{-1}$}
\newcommand{\cmt}   {cm$^{-3}$}
\newcommand{\cmd}   {cm$^{-2}$}
\newcommand{\jpb}   {$\rm Jy~beam^{-1}$} 
\newcommand{\lo}    {$L_{\sun}$}
\newcommand{\mo}    {$M_{\sun}$}
\newcommand{\mvir}  {$M_{\mathrm{vir}}$}
\newcommand{\nh}    {NH$_3$}

\newcommand{\hho}   {H$_2$O}
\newcommand{\hh}   {H$_2$}

\newcommand{\et}    {et al.}
\newcommand{\eg}    {e.\,g.,}

\newcommand{\vel} {$v_\mathrm{LSR}$}

\newcommand{\hii}	{\ion{H}{ii}}

\newcommand{\phnp}   {\phantom{0.}}
\newcommand{\phn}   {\phantom{0}}
\newcommand{\phnn}  {\phantom{0}\phantom{0}}

\newcommand{\phe}   {\phantom{$^\mathrm{c}$}}

\newcommand{\phb}   {\phantom{$>$}}

\begin{document}

\title{Low-mass protostars and dense cores in different evolutionary stages in IRAS\,~00213+6530}

\author{G. Busquet\inst{1}
        \and
	Aina Palau \inst{1,2}
	\and
        R. Estalella\inst{1}
        \and
        J. M. Girart\inst{3}
        \and
        G. Anglada\inst{4}
        \and
        I. Sep\'ulveda\inst{1}
         }

\offprints{Gemma Busquet,\\ \email{gbusquet@am.ub.es}}

\institute{Departament d'Astronomia i Meteorologia (IEEC-UB), Institut de Ci\`encies del Cosmos, Universitat de
	Barcelona, Mart\'{\i} i Franqu\`es 1, E-08028 Barcelona, Catalunya, Spain
         \and
	 Centro de Astrobiolog\'{\i}a (CSIC-INTA), Laboratorio de Astrof\'{\i}sica
	 Estelar y Exoplanetas, LAEFF campus, P.O. Box 78, E-28691 Villanueva de la
	 Ca\~nada (Madrid), Spain
	\and
	 Institut de Ci\`encies de l'Espai (CSIC-IEEC), Campus UAB,
         Facultat de Ci\`encies, Torre C-5 parell, E-08193 Bellaterra,
         Catalunya, Spain
         \and
	 Instituto de Astrof\'isica de Andaluc\'ia (CSIC), C/ Camino
	 Bajo de Hu\'etor 50, E-18008, Granada, Spain
	       }
 	 \date{Received / Accepted}

\authorrunning{G. Busquet \et}
\titlerunning{Star formation in IRAS\,~00213+6530}


\abstract{}{The aim of this paper is to study with high angular resolution a dense core associated with a low-luminosity IRAS source,
IRAS\,00213$+$6530, in order to investigate whether low mass star formation is really taking place in isolation.}{We carried out observations
at 1.2~mm with the IRAM\,30\,m telescope, and VLA observations in the continuum mode at 6~cm, 3.6~cm, 1.3~cm and 7~mm, together with \hho\
maser and \nh\ lines toward IRAS\,00213$+$6530. Additionally, we observed the CCS $J_N=2_1$--$1_0$ transition, and \hho\ maser emission using
the NASA 70~m antenna. We studied the nature of the centimeter and millimeter emission of the young stellar objects (YSOs) found in the
region, and the physical properties of the dense gas and dust emission.}{The centimeter and millimeter continuum emission, together with the
near infrared data from the 2MASS allowed us to identify three YSOs, IRS~1, VLA~8A, and VLA~8B, with different radio and infrared properties,
and which  seem to be in different evolutionary stages. IRS~1, detected only in the infrared, is in the more advanced stage. On the other
hand, VLA~8A, bright at centimeter and millimeter wavelengths, coincides with a near infrared 2MASS source, whereas VLA~8B has no infrared
emission associated with it and is in the earliest evolutionary stage. The overall structure of the NH$_3$ emission consists of three clouds.
Two of these, MM1 and MM2, are associated with dust emission at millimeter wavelengths, while the southern cloud is only detected in NH$_3$.
The YSOs are embedded in MM1, where we found evidence of line broadening and temperature enhancements. On the other hand, the southern cloud 
and MM2 appear to be quiescent and starless. Concerning the 1.2~mm dust emission, we modeled the radial intensity profile of MM1. The model
fits reasonably well the data, but it underestimates the intensity at small projected distances from the 1.2~mm peak, probably due to the
presence of multiple YSOs embedded in the dusty envelope. There is a strong differentiation  in the relative \nh\ abundance with low values
of $\sim2\times10^{-8}$ toward MM1, which harbors the YSOs, and high values, up to $10^{-6}$, toward the southern  cloud and MM2, suggesting
that these clouds could be in fact in a young evolutionary stage.}{IRAS\,00213$+$6530 is harboring a multiple system of low-mass protostars,
indicating that star formation in this cloud is taking place in groups or clusters, rather than in isolation. The low-mass YSOs found in
IRAS\,00213$+$6530 are in different evolutionary stages suggesting that star formation is taking place in different episodes.}

\keywords{
stars: formation --
ISM: individual objects: IRAS\,~00213+6530
-- ISM: clouds}

\maketitle

\section{Introduction}

It is widely accepted that there are two modes of star formation: the isolated mode and the clustered mode. This classification results from
studying the association between dense cores and young stellar objects (YSOs). For example, \citet{benson1989} study a wide sample of ammonia
cores and its relation with the position of IRAS sources, and find that typically only one IRAS source is associated with a single ammonia
core in the Taurus Molecular Cloud, where star formation can be assumed to take place in isolation. On the contrary Orion and Perseus would
be examples of molecular clouds forming stars in clustered mode \citep{lada1993}. A broad base of recent studies, carried out with higher
angular resolution than that of Benson \& Myers, show that most stars form in groups or clusters (\eg\ \citealt{clarke2000,lada2003}),
including low-mass stars (\eg\ \citealt{gomez1993,huard1999,lee2006,brooke2007,teixeira2007}), indicating that truly isolated star formation
is rare. \citet{adams2001} propose an intermediate case between the isolated and clustered modes, i.e., star formation in groups, and propose
that most stars form in groups and/or clusters, i.\,e., in cluster environments. However, the theories of low mass star formation assume that
star formation takes place in the isolated mode (\citealt{shu1987,lada1999}). Since star formation in cluster environments may differ from
the isolated mode (\eg\ \citealt{pfalzner2008}), it is necessary to study with high angular resolution dense cores associated with one single
IRAS source to assess if star formation is really taking place in isolation. In this context, we aim at investigating with high angular
resolution a dense ammonia core associated with a single low luminosity IRAS source, IRAS\,~00213+6530 (hereafter I00213).

\begin{small}
\begin{table*}[t]
\caption{VLA observational parameters in the IRAS\,~00213+6530 region
\label{tjcmt}}
\begin{tabular}{lcccccccccc}
\hline\hline
&&\multicolumn{2}{c}{Phase Center}
&&Bootstrapped
&\multicolumn{2}{c}{Synthesized Beam}&&On-Source\\
\cline{3-4}
\cline{7-8}
&$\lambda$& $\alpha$(J2000)& $\delta$(J2000)& Phase& Flux Density& HPBW& P.A &rms Noise &Time\\
Transition & $\mathrm{(cm)}$ &(h m s)& ($\degr$ $'$ $''$)&Calibrator &(Jy)& (arcsec)&(deg)& (m\jpb)&(hours)& Epoch\\ 
\hline
continuum& \phnp6    &00 24 10.41& $+$65 47 02.0&  0014$+$612&  $1.940\pm0.006$&	 17.9\,$\times$11.7&             \phb54&   0.024\phe& 	     1.2 &2000 \\
continuum&   3.6     &00 24 10.41& $+$65 47 02.0&  0014$+$612&  $1.281\pm0.010$&	 21.2$\times$\phn9.2&           $-68$&   0.023\phe&   1.0 &2000 \\
continuum&   3.6     &00 24 11.44& $+$65 47 09.6&  0019$+$734&  $1.070\pm0.003$&   	 \phn9.9\,$\times$\phn7.2& \phb55&   0.027\phe&         0.5 &2004\\ 
continuum&   1.3     &00 24 11.44& $+$65 47 09.6&  0019$+$734&  $2.41\phn\pm0.08\phn$&	 \phn1.5\,$\times$\phn1.0&       \phb32&  0.15\phn\phe&  0.3& 	2006\\
continuum&   0.7     &00 24 11.44& $+$65 47 09.6&  0019$+$734&  $1.71\phn\pm0.06\phn$&   \phn2.9\,$\times$\phn2.6&  $-$89&   0.19\phn\phe&  1.0 &2004\\
\nh\,$(1,1)$& 1.3    &00 24 11.44& $+$65 47 09.6&  0019$+$734&  $1.58\phn\pm0.05\phn$&   \phn4.1\,$\times$\phn3.7&  $-$52&   1.3$^\mathrm{a}$\phnn&  \phnp6  &2004\\
\nh\,$(2,2)$& 1.3    &00 24 11.44& $+$65 47 09.6&  0019$+$734&  $1.58\phn\pm0.05\phn$&   \phn4.1\,$\times$\phn3.7&  $-$51&   1.1$^\mathrm{a}$\phnn&  \phnp6 &2004 \\
\hho\ 6$_{16}$--$5_{23}$&         1.3    &00 24 11.44& $+$65 47 09.6&  0019$+$734&  $2.41\phn\pm0.08\phn$&   \phn2.3\,$\times$\phn1.4& \phb17&   3.0$^\mathrm{a}$\phnn&  0.3 &2006\\
\hline
\end{tabular}
\begin{list}{}{}
\item[$^\mathrm{a}$] per channel
\end{list}
\label{tvlaobs}
\end{table*}
\end{small}

I00213, with a luminosity of $\lesssim20$~\lo\ and at 850~pc of distance, belongs to the molecular cloud M120.1+3.0 \citep{yang1990} in
the Cepheus\,OB4 star-forming region.  The region is physically related to the \hii\ region S171 \citep{yang1990}.  The \nh\ emission in
the north of M120.1+3.0 was studied through single-dish observations by \citet{sepulveda2001}. The \nh\ emission consists of two
condensations, each one peaking very close to the position of an IRAS source, I00213 and IRAS~00217+6533 (I00217), suggesting
that both IRAS sources are deeply embedded in high density gas. The mass derived for the condensation associated with I00213 is
$\geq45$~\mo.  The ammonia emission engulfing both IRAS sources is associated with CO high-velocity emission, indicating the presence of
a molecular outflow in the region (\citealt{yang1990}). However, it is not clear which IRAS source is driving the outflow.


In this paper we report on high angular resolution observations with the Very Large Array (VLA) of the continuum
emission at 6~cm, 3.6~cm, 1.3~cm, and 7~mm, as well as of the dense gas traced by \nh\,(1,1) and \nh\,(2,2) together
with observations of \hho\ maser emission. In addition we also present the continuum emission at 1.2~mm observed with
the IRAM\,30~m telescope, and CCS and \hho\ maser observations carried out with the NASA\,70\,m antenna at Robledo de
Chavela.  The paper layout is the following: in \S\,2 we describe our observations and the data reduction process,
and present the main results for the continuum and molecular line emission in \S\,3. In \S\,4 we analyze the dust and
\nh\ emission and show the method used to derive the \nh\ abundance in this region. Finally, in \S\,5 we discuss our
findings, and we list the main conclusions in \S\,6.


\section{Observations}

\subsection{IRAM\,30~m observations}

The MPIfR 37-element bolometer array MAMBO at the IRAM\,30~m telescope\footnote{IRAM is supported by INSU/CNRS (France), MPG (Germany), and
IGN (Spain)} was used to map the 1.2~mm dust continuum emission toward I00213. The observations were carried out in 2006 June 2.  The main
beam has a HPBW of 10$''$. We used the on-the-fly mapping mode, in which the telescope scans continuously in azimuth along each row. The
sampled area was $200''\times140''$, and the scanning speed was 5$''$~$\rm {s}^{-1}$. Each scan was separated by 5$''$ in elevation. The
secondary mirror was wobbling at a rate of 2~Hz in azimuth with a wobbler throw of 46$''$.The average zenith opacity was in the range
0.3--0.4. Pointing and focus were done on NGC\,7538. The rms of the final map was $\sim\!3.8~$m\jpb. Data reduction was performed with the
MOPSIC\footnote{See http://www.iram.es/IRAMES/mainWiki/CookbookMopsic} software package that contains the necessary scripts for data
reduction (distributed by R. Zylcka).

\subsection{VLA radio continuum observations} 

The observations were carried out using the VLA of the NRAO\footnote{The Very Large Array (VLA) is operated by the
National Radio Astronomy Observatory (NRAO), a facility of the National Science Foundation operated under cooperative
agreement by Associated Universities, Inc.} in the D configuration in the continuum mode at 6~cm and 3.6~cm on 2000
September 23, and at 3.6~cm and 7~mm during 2004 August 24. The observational parameters for each epoch are
summarized in Table~\ref{tvlaobs}. During the first epoch absolute flux calibration was achieved by observing 3C286,
with an adopted flux density of 7.49~Jy at 6~cm and 5.18~Jy at 3.6~cm. The absolute flux calibrator was 00137+331
(3C48) during the 2004 observations, for which flux densities of 3.15 Jy and 0.53 Jy were adopted at 3.6~cm
and 7~mm, respectively. In order to minimize the effects of atmospheric phase fluctuations, at 7~mm we used
the fast switching technique \citep{carilli1997} between the source and the phase calibrator over a cycle of 80
seconds, with 50 seconds spent on the source and 30 seconds on the calibrator. The 1.3~cm continuum emission was
observed on 2006 December 2 simultaneously with the \hho\ maser emission (see below, Sect.~2.3).


Calibration and data reduction were performed using standard procedures of the Astronomical Imaging Processing System (AIPS)\footnote{See
http://aips.nrao.edu} of the NRAO. Clean maps at 3.6~cm and 7~mm were made using the task IMAGR of AIPS with the robust parameter of
\citet{briggs1995} set equal to 5, which is close to natural weighting, whereas the map at 6~cm was made with the robust parameter equal to
zero in order to obtain a synthesized beam similar to that at 3.6~cm. Since the signal-to-noise ratio of the longest baselines of the 7~mm
data was low, we applied a $uv$-taper function of 80~k$\lambda$ in order to improve the sensitivity.

\subsection{VLA \nh\ and \hho\ maser observations}

The observations of $(J,K)=(1,1)$ and $(J,K)=(2,2)$ inversion lines of the ammonia molecule were carried out in the same run as
the 2004 continuum observations. In Table~\ref{tvlaobs} we summarize the observational parameters.  The adopted flux density of
the absolute flux calibrator 0137+331 (3C48) was 1.05~Jy at a wavelength of 1.3~cm, and the bandpass calibrator used was 0319+415
(3C84). We used the 4IF spectral line mode, which allows simultaneous observations of the \nh\,$(1,1)$ and $(2,2)$ lines with two
polarizations for each line. The bandwidth used was 3.12~MHz, with 63 channels with a channel spacing of 48.8~kHz
(0.6~\kms\ at 1.3~cm) centered at $v_\mathrm{LSR}=-$19.0~\kms, plus a continuum channel that contains the average of the central
75~\%\ of the bandwidth. 


The water maser line at 22.2351~GHz (6$_{16}$--$5_{23}$ transition) was observed with the VLA in the C configuration during 2006
December 2. The phase center was the same as for the \nh\ observations, and the adopted flux density of the absolute flux calibrator,
0137+331 (3C48), was 1.13~Jy at 1.3~cm. We summarized the observational parameters in Table~\ref{tvlaobs}. We used the 4IF
mode, employing two IF with a total bandwidth of 3.12~MHz, with 63 channels with a channel spacing of 48.8~kHz (0.6~\kms) centered at
$-$10~\kms\ \footnote{Note that the center velocity for the \hho\ maser line observed with the VLA is shifted by 9~\kms\ from that of
the \nh\ observations since the \hho\ maser reported by \citet{han1998} was detected at $v_\mathrm{LSR}=-$0.7~\kms}, and  two IF with a
total bandwidth of 25~MHz were used to observe the continuum emission.

The \nh\ and \hho\ data were reduced with the standard AIPS procedures. The images were constructed using natural weighting in both cases.


\subsection{NASA\,70\,m CCS and \hho\ maser observations}

We carried out a \hho\ maser emission monitoring toward IRAS\,00213$+$6530 with the NASA\,70\,m antenna (DSS-63) at
Robledo de Chavela (Spain).  The observations were performed in 2008 Apr 18, June 19, and September 23 using a cooled
high-electron-mobility transistor as 1.3~cm front-end, and a 384 channel spectrometer as backend, covering a bandwidth of 16~MHz
($\sim 216$~\kms\ with 0.6~\kms\ resolution). Spectra were taken in position-switching mode. The HPBW of the telescope at this
frequency is $\sim 41''$. The typical system temperature was 120 K and the total integration time was around 30 min (on+off) for
each session. 

In addition to the \hho\ maser observations, we also observed the CCS $J_N=2_1$--$1_0$ transition (22.344~GHz), with the same bandwidth and
spectral resolution used for \hho. The CCS transition was observed on October 3rd 2008 and on February 7th 2009 during a total integration
time of 26 and 40 minutes, respectively. The system temperature was 60~K and 73~K, respectively.

For all the observations, the rms pointing accuracy of the telescope was better than $10''$. A noise diode was used to calibrate the data,
and the uncertainty in the flux calibration is estimated to be $\sim30$~\%. The data reduction was performed using the CLASS package, which
is part of the GILDAS\footnote{See http://www.iram.fr/IRAMFR/GILDAS} software.



\begin{table}[t]
\caption{Parameters of the 1.2~mm emission
\label{tjcmt}}
\begin{tabular}{lcccc}
\hline\hline
&\multicolumn{2}{c}{Position}&&\\
\cline{2-3} 
 &$\alpha$ (J2000) &$\delta$ (J2000)&$I_{\nu}$$^{{\mathrm{peak}}}$ &$S_{\nu}$  \\ 
 & (h m s)& ($\degr$ $'$ $''$)& (m\jpb)& (mJy) \\
\hline
MM1&  00 24 10.7& 65 47 09& 146& 332$\pm$35  \\
MM2&  00 24 04.4& 65 48 12& 46& 146$\pm$\phn9  \\
\hline
\end{tabular}
\label{iram30m}
\end{table}

\section{Results}

\subsection{Continuum at 1.2~mm}

\begin{figure}[t]
\begin{center}
\begin{tabular}[b]{c}
	\epsfig{file=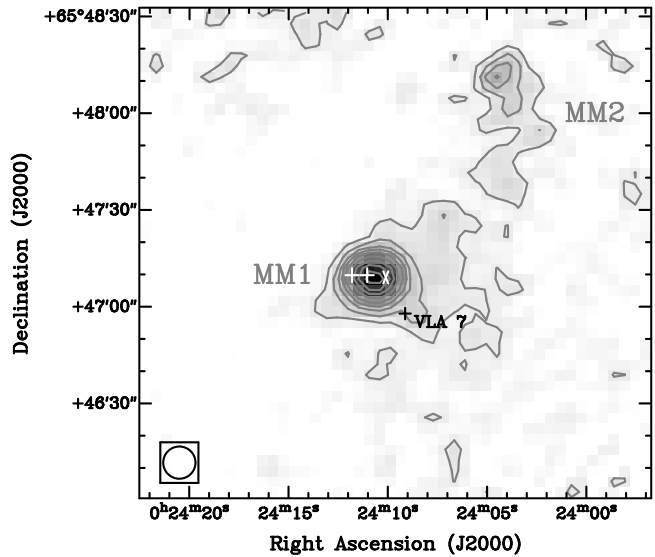,scale=0.70} \\
       \end{tabular}
      \caption{1.2~mm continuum emission toward IRAS~00213+6530.
       Contour levels are 3, 6, 9, 12, 15, 20, 25, 30 and 35 times the rms of the map, 3.8~m\jpb. The
       synthesized beam of the IRAM\,30\,m is shown in the bottom left corner of the image. White
       crosses  indicate the position of the millimeter sources VLA~8A and VLA~8B, and the white tilted
       cross marks the position of IRS~1. VLA~7 is indicated by the black cross (see
       Sect.~3.2 for the objects labeled in this figure).}
\label{firam30m}
\end{center}
\end{figure}

Figure~\ref{firam30m} shows the 1.2~mm continuum emission  observed with the IRAM\,30~m telescope toward I00213. The overall structure of
the dust emission consists of a central and compact dust condensation, MM1, with some extended structure to the west. A 2D Gaussian fit to
MM1 yields a deconvolved size of $13''\times12''$, P.A.$=92\degr$. In addition, we also detected a fainter dust condensation, MM2, located
to the northwest of MM1, and elongated in the north-south direction, in a filamentary structure connecting MM1 and MM2. The
main results are summarized in Table~\ref{iram30m}. 



\subsection{VLA radio continuum emission}

\begin{table*}[!t]
\caption{Parameters of the continuum sources detected in the IRAS 00213+6530 region
\label{tjcmt}}
\begin{center}
\begin{tabular}{lccccc}
\hline\hline
&\multicolumn{2}{c}{Position$^\mathrm{a}$}&	Flux Density&	Flux
Density&	Spectral\\
\cline{2-3}
&$\alpha$ (J2000) &$\delta$ (J2000) &at 6~cm &at 3.6~cm &Index\\
Source&	(h m s)& ($\degr$ $'$ $''$)& (mJy)& (mJy)& 6~cm--3.6~cm \\
\hline
1&  00 23 21.01& $+$65 43 46.5&  0.87$\pm$0.07& \ldots$^\mathrm{b}$&             \ldots\\
2&  00 23 45.65& $+$65 48 36.1&  0.17$\pm$0.03& $<$0.18$^\mathrm{c}$&         $<$0.10\\
3&  00 23 46.41& $+$65 49 12.7&  1.66$\pm$0.04& 0.38$\pm$0.07$^\mathrm{d}$&   $-$2.7$\pm$0.3\\
4&  00 23 49.01& $+$65 46 14.7&  0.69$\pm$0.03& 0.42$\pm$0.02&                $-$0.9$\pm$0.1\\
5&  00 23 52.06& $+$65 39 58.9&  4.20$\pm$0.60& \ldots$^\mathrm{b}$&             \ldots\\
6&  00 23 53.70& $+$65 49 41.6&  0.86$\pm$0.03& $<$0.25$^\mathrm{c, d}$&      $<-2.1$\\
7&  00 24 09.11& $+$65 46 58.0&  0.57$\pm$0.03& 0.24$\pm$0.02&                $-$1.6$\pm$0.2\\
8&  00 24 11.39& $+$65 47 09.6&  0.16$\pm$0.03& 0.21$\pm$0.02&                \phb0.5$\pm$0.4\\
9&  00 24 24.46& $+$65 49 05.9&  0.13$\pm$0.03& 0.27$\pm$0.02&                \phb1.3$\pm$0.4\\
10& 00 24 36.71& $+$65 48 27.5&  0.55$\pm$0.03& 0.22$\pm$0.03&                $-$1.7$\pm$0.3\\
11& 00 24 56.70& $+$65 49 10.3&  0.63$\pm$0.05& \ldots$^\mathrm{b}$&             \ldots\\
12& 00 25 24.75& $+$65 45 48.3&  4.90$\pm$1.10& \ldots$^\mathrm{b}$&             \ldots\\
\hline
\end{tabular}
\end{center}
\begin{list}{}{}
\item[$^\mathrm{a}$] Positions taken from the 3.6~cm map, except for
non-detected sources at this band, for which positions correspond to
the 6~cm map.
\item[$^\mathrm{b}$] Source well outside the primary beam at this wavelength. Primary beam correction larger than
12, and the uncertainty in the corrected flux is very large.
\item[$^\mathrm{c}$] The upper limit for non-detected sources is 4$\sigma$.
\item[$^\mathrm{d}$] Flux density measured in 2000 September 23. Source 
highly variable, so to compute the spectral index we used
observations carried out at the same epoch.
\end{list}
\label{tvlapar}
\end{table*}

\begin{figure*}[t]
\begin{center}
\begin{tabular}{cc}
    \epsfig{file=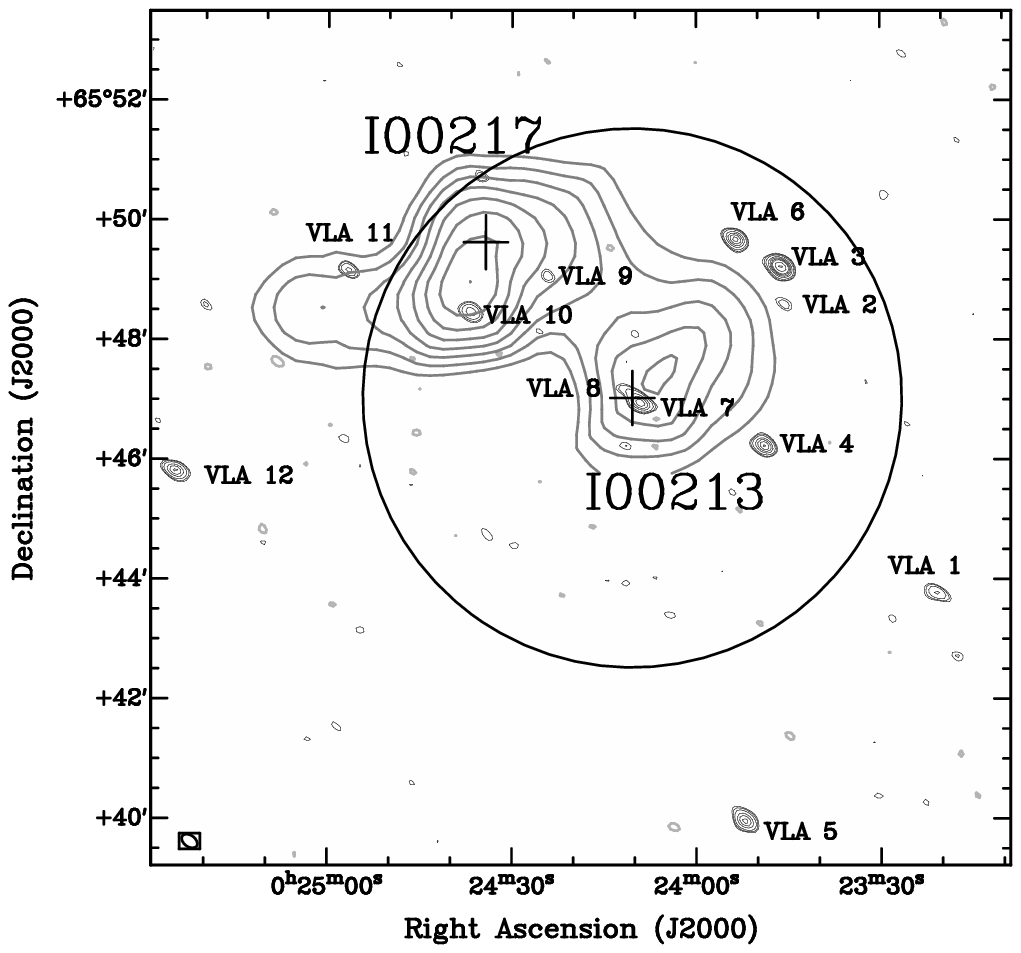,scale=0.85}&
     \epsfig{file=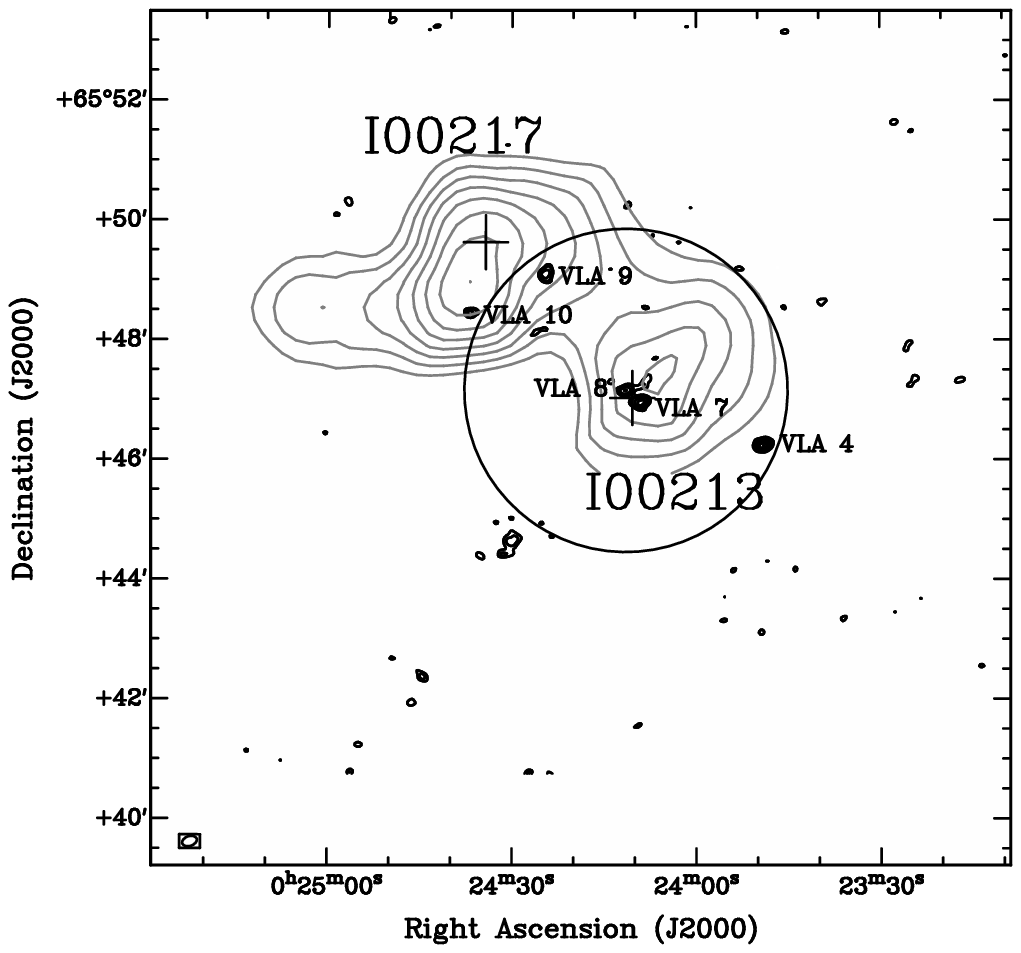,scale=0.85}
    \end{tabular}
 
    \caption{In both panels grey contours represent the main beam brightness temperature of the main line of
the \nh\ $(J,K)=(1,1)$ inversion transition from Sep\'ulveda (2001). \emph{Left:} VLA  6~cm continuum
emission map (black contours) of the I00213 region.  Contour levels are $-$3, 3, 4, 6, 10, 14, 18, 22, 32, and 42
times the rms of the map, 24~$\mu$\jpb. The synthesized beam, $17\farcs9\times11\farcs7$, with P.A.$= 54\degr$, is
shown in the bottom left corner of the image. \emph{Right:} VLA 3.6~cm continuum emission map (black
contours) of the I00213 region. Contour levels are $-$3, 3, 4, 6, 8, 10, and 12 times the rms of the map,
18~$\mu$\jpb. The synthesized beam, $15\farcs14\times9\farcs89$, with P.A.$=-72\degr$, is shown in the bottom left
corner of the image. The positions of the IRAS sources are indicated by crosses. IRAS\,00213$+$6530 is at
the center of the figure, while IRAS\,00217$+$6533 is at the north-east. The VLA primary beam, 9$\arcmin$ at 6~cm and
5$\arcmin$ at 3.6~cm are also indicated by black circumferences.}

\label{vlacm}
\end{center}
\end{figure*}

\begin{figure}[ht]
\begin{center}
\begin{tabular}{c}
	\epsfig{file=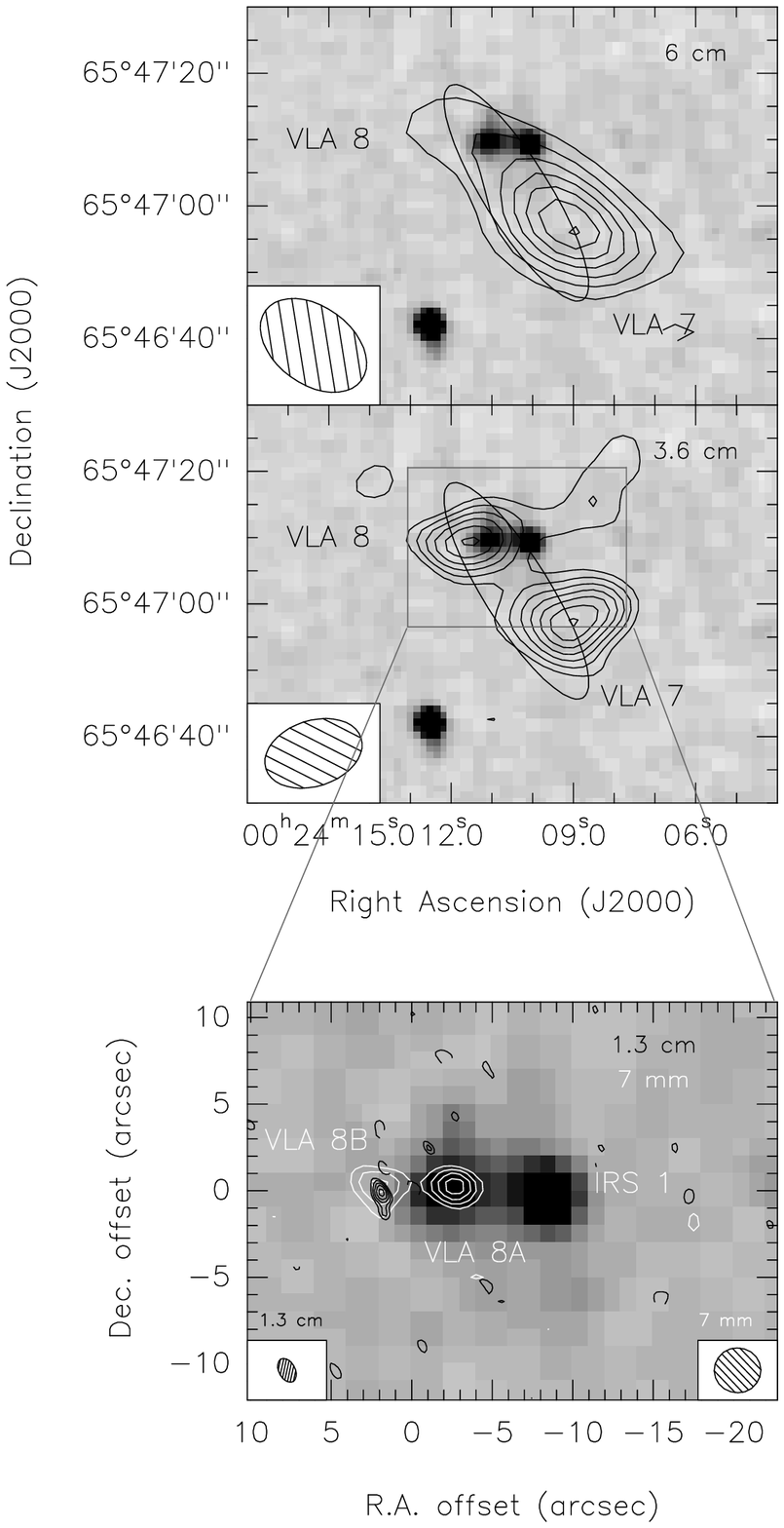, scale=0.7}
    \end{tabular}

     \caption{\emph{Top:} VLA contour map of the 6~cm continuum emission toward I00213. Contour levels are $-$3, 3, 6, 9, 12, 15, 18, and 21
times the rms of the map, 24~$\mu$\jpb. \emph{Middle:} VLA 3.6~cm continuum emission. Contour levels are -3, 3, 4, 5, 6, 7, 8 and 9
times the rms of the map, 18~$\mu$\jpb. The synthesized beams at 6~cm ($17\farcs9\times11\farcs7$, with P.A.$=54\degr$), and at 3.6~cm
($15\farcs1\times9\farcs9$, with P.A.$= -72\degr$), are shown at the bottom left corner of each panel. \emph{Bottom:} White: VLA 7~mm
continuum emission, black: VLA 1.3~cm continuum emission. Contour levels at 7~mm are $-$3, 3, 4, 5, and 6 times the rms of
the map, 0.2~m\jpb. At 1.3~cm contour levels are $-$3, 3, 4, 5, 6, and 7 times the rms of the map, 0.15~m\jpb. The synthesized beams at
1.3~cm and 7~mm are shown at the bottom left and right corners of the image, respectively. In all panels grey scale is the 2MASS
$K_\mathrm{s}$-band image, and the position error of IRAS~00213+6530 is indicated by the ellipse.} 

\label{fvla78} 
\end{center}
\end{figure}

We detected 11 sources at 6~cm, and 7 sources at 3.6~cm above the 4~$\sigma$ detection threshold.  Figure~\ref{vlacm} shows the 6~cm and
3.6~cm continuum emission observed with the VLA toward I00213.  In Table~\ref{tvlapar} we list the positions and flux densities, corrected
for primary beam response, of the detected sources, and the estimated spectral indices between 6~cm and 3.6~cm. To obtain the 3.6~cm map we
analyzed separately the observations of the two epochs (see Sect.~2.2) in order to see the degree of variability of the detected sources at
this wavelength. The final map was obtained after subtracting the $uv$ data of VLA~3, which presents a variability larger than 8$\sigma$, and
VLA~6, detected in 2004 but not during the observations carried out in 2000. In Table~\ref{tvar} we show the flux density measured in 2000
and 2004 for VLA~3 and VLA~6, as well as the variability during this period. Both sources, VLA~3 and VLA~6, have a negative spectral index
(from simultaneous observations at 6~cm and 3.6~cm) and are probably non-thermal extragalactic background sources.

As can be seen in Fig.~\ref{fvla78} (middle panel), at 3.6~cm we detected two sources toward I00213, VLA~7 and VLA~8, separated by
$20"$ and both inside the position error ellipse of the IRAS source. These sources are barely resolved at 6~cm, with VLA~8 just
being a weak prolongation to the north-east of VLA~7 (see Fig.~\ref{fvla78} top panel). Yet, the higher angular resolution of the
3.6~cm map allows to separate the two sources. VLA~8 peaks close to the position of the dust condensation MM1, whereas VLA~7 lies
$\sim20''$ to the south-west, in the extended structure of the dust emission (see Fig.~\ref{firam30m}). Both sources are spatially
resolved and VLA~8 shows a weak tail extending to the west. A 2D Gaussian fit to the two sources (excluding the weak tail of VLA~8)
yields deconvolved sizes of $9\farcs7\times2\farcs2$ (P.A.$=46\degr$), and $13\farcs4\times3\farcs1$ (P.A.$=63\degr$) for VLA~7 and
VLA~8, respectively.  The spectral index in the 6--3.6~cm range of VLA~7 is $-$1.6$\pm$0.2, characteristic of non-thermal emission,
whereas VLA~8 has a spectral index of 0.5$\pm$0.4, which is consistent with free-free thermal emission from ionized gas that may be
arising from a thermal radio jet. Positive spectral indices, i.\,e., $\alpha\gtrsim -0.1$, have been found to be associated with
sources driving molecular outflows (\eg\ \citealt{anglada1998,beltran2001}). The 2MASS $K_\mathrm{s}$-band image shows two sources,
2MASS J00241110+6547095 and 2MASS J00241010+6547091, the first nearly coinciding with VLA~8 and the second, named IRS~1, lying
$\sim6''$ to the west.

\begin{table}[t]
\caption{Highly variable sources at 3.6~cm
\label{tjcmt}}
\begin{tabular}{lccc}
\hline\hline
&2000 Obs.& 2004 Obs.&  Variability\\
& $S_{\nu}\,$(3.6 cm)& $S_{\nu}\,$(3.6 cm)& with respect\\
Source& (mJy)& (mJy)&   to 2004\\
\hline
VLA 3&  $0.38\pm0.07$&   	$1.86\pm0.07$&	\phb79\%    ($\sim21\sigma$)\\
VLA 6&  $<0.25^\mathrm{a}$&		$0.41\pm0.05$&  $>$39\%     ($>\phn3\sigma$)\\
\hline
\end{tabular}
\begin{list}{}{}
\item[$^\mathrm{a}$] The upper limit for non-detected sources is 4$\sigma$.
\end{list}
\label{tvar}
\end{table}





The maps of the 1.3~cm and 7~mm continuum emission obtained with natural weighting are shown in Fig.~\ref{fvla78} (bottom panel). While we
did not detect 7~mm continuum emission toward VLA~7, the 7~mm emission of VLA~8 is resolved into two components, VLA~8A and VLA~8B, separated
by $\sim5''$ ($\sim$~4300~AU at the distance of the source). At 1.3~cm we detected one source associated with VLA~8B, whose peak position
coincides within $\sim0\farcs$5 with the 7~mm peak (Fig.~\ref{fvla78} bottom panel). The 1.3~cm source is elongated roughly in the
northeast-southwest direction, and is spatially resolved only in one direction, with a deconvolved size of $1\farcs2$ ($\sim1020$~AU), at
P.A.$=21\degr$. In Table~\ref{tvla7mm} we show the position, peak intensity, and flux density of VLA~8A and VLA~8B at 1.3~cm and 7~mm, as
well as the spectral index between these wavelengths. In order to properly estimate the spectral index in the 1.3~cm--7~mm range we applied a
uv-taper function of 80~k$\lambda$ to obtain similar angular resolutions at both wavelengths (see Sect.~2.2). The resulting spectral indices
are $>1.4$ and 0.2$\pm$0.6 for VLA~8A and VLA~8B, respectively.



\begin{table*}[t]
\caption{Parameters of sources detected at 1.3~cm and 7~mm in the IRAS 00213+6530 region
\label{tjcmt}}
\begin{center}
\begin{tabular}{lccccccccc}
\hline\hline
&\multicolumn{2}{c}{Position$^\mathrm{a}$} 
&&\multicolumn{2}{c}{1.3~cm}
&&\multicolumn{2}{c}{7~mm}
&Spectral \\
\cline{2-3}
\cline{5-6}
\cline{8-9}
&$\alpha$ (J2000) &$\delta$ (J2000) & &$I_{\nu}$$^{{\mathrm{peak}}}$ &$S_{\nu}$
&&$I_{\nu}$$^{{\mathrm{peak}}}$ &$S_{\nu}$& Index$^\mathrm{c}$\\
Source & (h m s)  & ($\degr$ $'$ $''$)&&(m\jpb)& (mJy)&& (m\jpb)& (mJy)& 1.3~cm-7mm\\
\hline
VLA~8A&  00 24 11.01& 65 47 09.9& & \ldots&  $<0.6$$^\mathrm{b}$&  &1.22$\pm$0.26& 1.57$\pm$0.54& $>1.4$ \\
VLA~8B&  00 24 11.76& 65 47 09.8& & 1.09$\pm$0.13& 1.25$\pm$0.25& & 1.12$\pm$0.38& 1.43$\pm$0.79 &0.2$\pm$0.6 \\
\hline
\end{tabular}
\label{tvla7mm}
\end{center}
\begin{list}{}{}
\item[$^\mathrm{a}$] From the 7~mm map.
\item[$^\mathrm{b}$] 4~$\sigma$ upper limit.
\item[$^\mathrm{c}$] Estimated from the 1.3~cm flux density obtained with a uv-taper
function of 80~k$\lambda$ (see main text)
\end{list}
\end{table*}

\begin{figure*}[t]
\begin{center}
\begin{tabular}{c}
    \epsfig{file=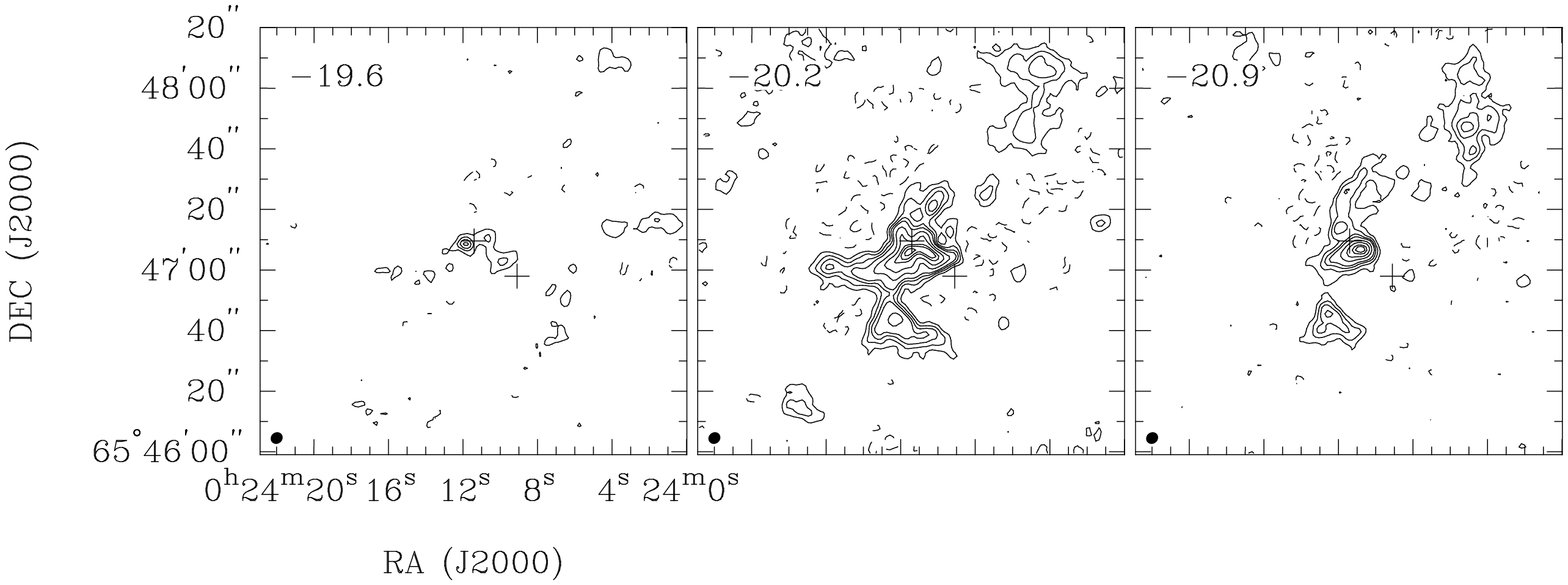,scale=0.7,angle=-90}\\
    \epsfig{file=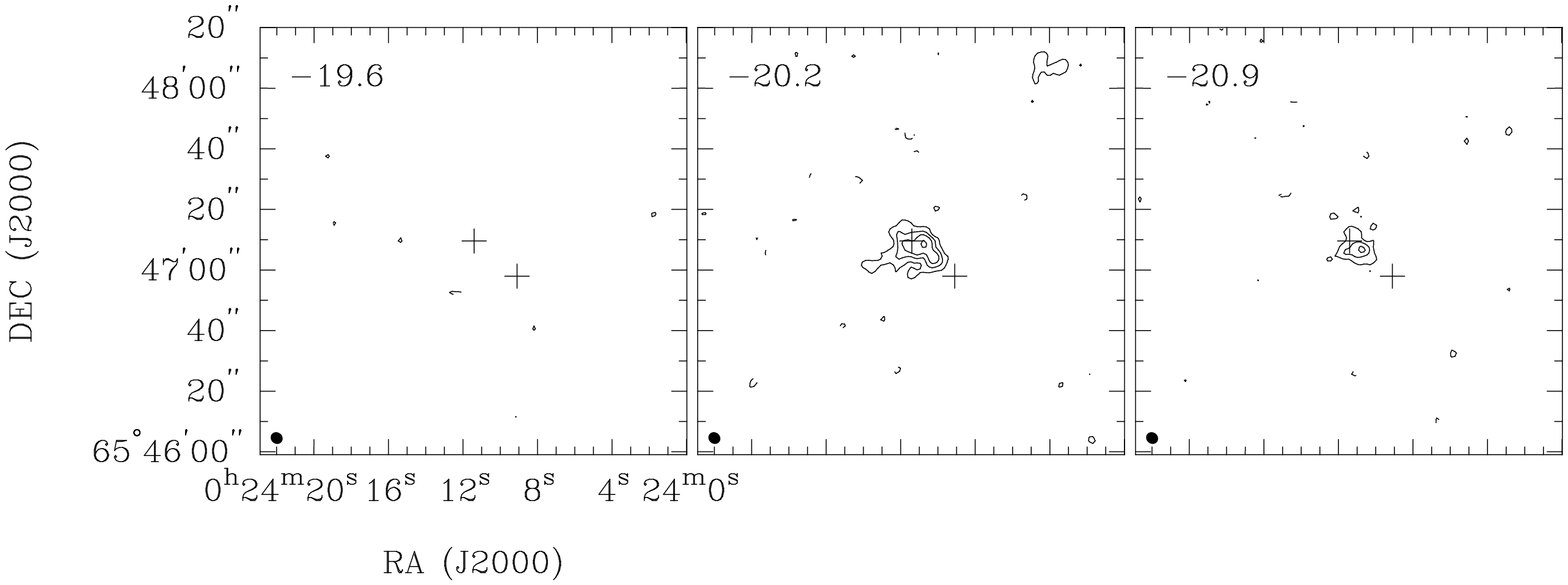,scale=0.7,angle=-90} 
    \end{tabular}
     \caption{\emph{Top panel:} VLA channel maps of the \nh\,(1,1)
main line. Contours levels are $-$6, $-$3, 3, 6, 9, 12, 18, 24, 30, 33, and 36
times the rms noise of the map, 1.1~m\jpb.  \emph{Bottom panel:} VLA channel maps of the
\nh\,(2,2) main line. Contour levels are $-3$, 3, 6, 9, and 12 times
the rms noise of the map, 1.1~m\jpb . In both panels the synthesized beam is shown in the bottom
left corner. The positions of VLA~7 and VLA~8 are indicated by
crosses.}
\label{fnh3ch}
\end{center}
\end{figure*}

\begin{figure}[t]
\begin{center}
\begin{tabular}[b]{cc}
    \epsfig{file=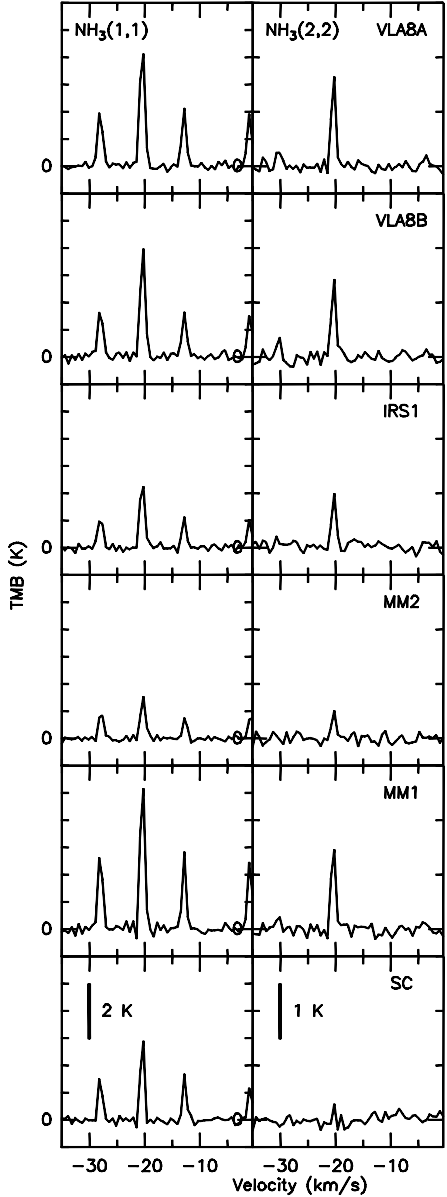,scale=0.68} 
    \end{tabular}

 \caption{Spectra toward six positions of the IRAS\,00213+6530 region for \nh\,$(1,1)$ (\emph{left}) and
	  \nh\,$(2,2)$ (\emph{right}), averaged over one beam. The six positions, which are
	  labeled on the right panel of each row, are, from top to bottom, VLA~8A, VLA~8B, IRS~1, MM2 (peak of the
	  northwestern cloud), MM1 (peak of the central cloud), and SC (peak of the southern cloud). The vertical
	  scale for each transition is indicated in the bottom row.}

\label{fnh3spec}
\end{center}
\end{figure}

\begin{figure}[t]
\begin{center}
\begin{tabular}[b]{c}
    \epsfig{file=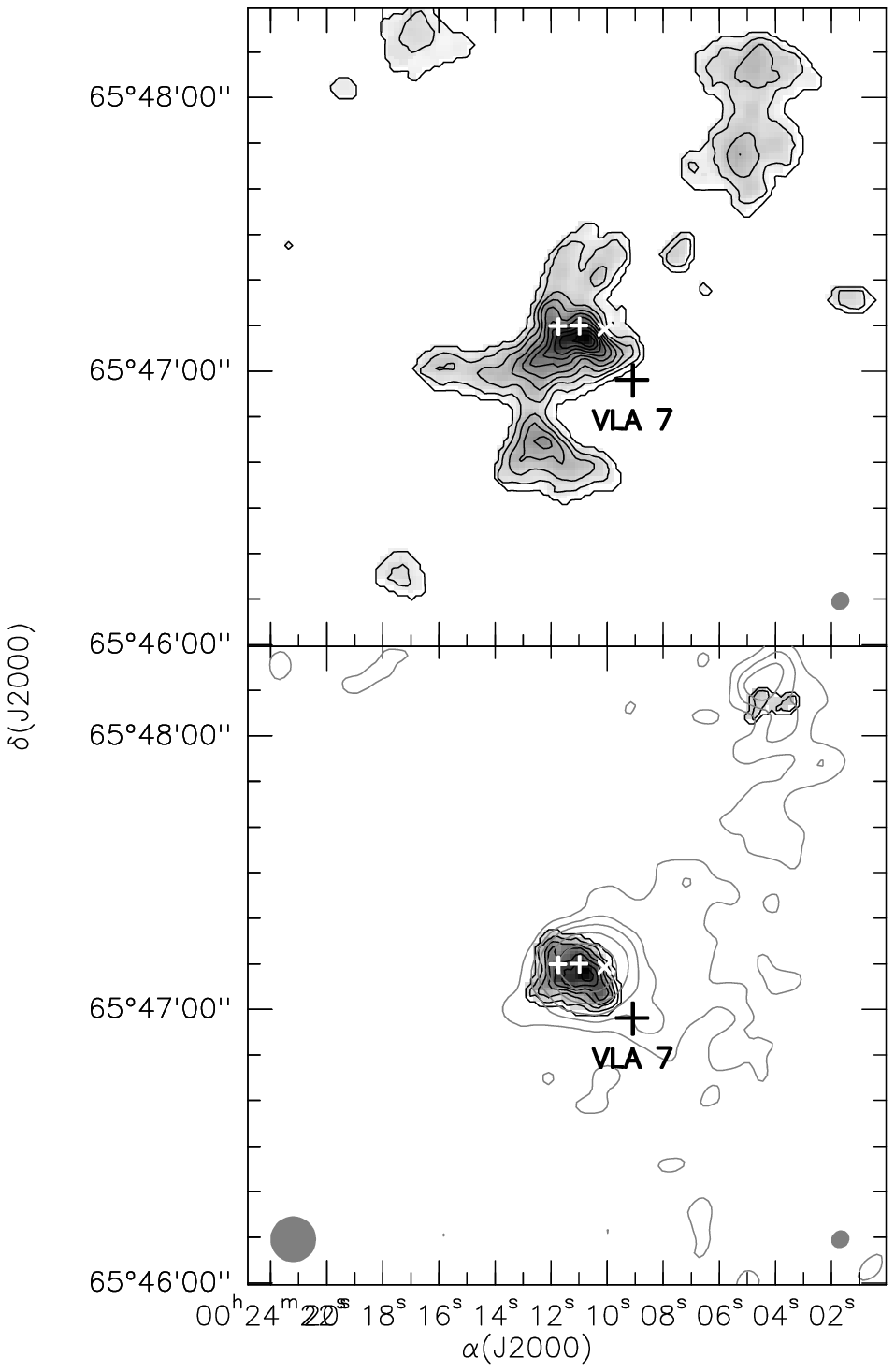,scale=0.8} \\
    \end{tabular}

       \caption{\emph{Top panel:} Zero-order moment of the \nh\,$(1,1)$ main line emission. Contours
    start at 1~\%, increasing in steps of 10~\% of the peak intensity, 0.0508~\jpb~\kms. \emph{Bottom panel:}
    Zero-order moment map of the \nh\,$(2,2)$ main line emission (black contours). Contours start at 1~\%,
    increasing in steps of 15~\% of the peak intensity, 0.0151~\jpb~\kms. Grey contours: 1.2~mm continuum
    emission. Contour levels are 3, 6, 9, 12, 25, and 35 times the rms of the map, 3.8~m\jpb. In both panels white
    crosses mark the position of the two millimeter sources, VLA~8A and VLA~8B, and the white tilted cross marks
    the position of the infrared source IRS~1. The synthesized beams for each transition are shown at the bottom
    right corner, and the synthesized beam of the 1.2~mm continuum data is shown in the bottom left corner of the
    bottom panel.}

\label{fnh3mom0}
\end{center}
\end{figure}

As seen in Fig.~\ref{fvla78} (bottom panel), VLA~8A coincides with the near infrared source 2MASS J00241110+6547095, whereas VLA~8B has no
infrared emission associated with it. In addition, $\sim6''$ ($\sim$5100~AU ) west of VLA~8A there is the near-infrared source IRS~1 with no
detected 7~mm emission. At 7~mm both thermal dust emission and free-free emission can contribute to the total emission. Then, in order to
estimate the mass of gas and dust from the 7~mm emission, we need to know first the flux density coming from thermal dust emission. Since the
angular resolution achieved at 3.6~cm is not sufficient to resolve the two millimeter sources we cannot estimate the contribution of
free-free emission at 7~mm for each individual source. A first approach is to smooth the 7~mm emission to the angular resolution of the
3.6~cm data. The result is a single source with a flux density of 2.5\,$\pm$\,0.9~mJy, which is consistent with the sum of the flux densities
of both millimeter sources. Extrapolating the flux obtained at 3.6~cm for VLA~8 to millimeter wavelengths with the spectral index obtained
from the centimeter emission at 6~cm and 3.6~cm ($\alpha$=0.5), we find that the expected free-free emission at 7~mm is $\sim0.48$~mJy
($19~\%$ of the 7~mm flux), indicating that the thermal dust component dominates at this wavelength. We note that the free-free contribution
estimated at 7~mm can be considered as an  upper limit because the free-free spectral index between 3.6~cm and 7~mm is expected to be flatter
than between 6 and 3.6~cm. However, the thermal dust contribution at 7~mm may be different for VLA~8A and VLA~8B. The fact that VLA~8A is not
detected at 1.3~cm above a 4~$\sigma$ level indicates that at 7~mm the emission is mainly due to thermal dust emission rather than free-free
emission, whereas VLA~8B, which is associated with a 1.3~cm source, must have less thermal dust emission associated with it.  

To estimate the mass, we assumed that the dust emission is optically thin, and used the opacity law of
$\kappa_\mathrm{{\nu}}=0.01(\nu/230~\mathrm{GHz)}^{\beta}$~$\mathrm{cm^{2}~g^{-1}}$ \citep{ossenkopf94}, extrapolated to 7~mm. We used a
dust emissivity index $\beta=1.5$ (derived from the spectral energy distribution, see Sect.~4.2). The dust temperature is estimated by
correcting the rotational temperature derived from \nh\  ($\sim 20$~K, see next sections) to kinetic temperature ($\sim 25$~K),
following the expression of \citet{tafalla2004}.  Using the fraction of the 7~mm flux density arising from thermal dust emission
($\sim2.1$~mJy), the total mass derived for the two sources, VLA~8A and VLA~8B, is 6.3~\mo. It is worth noting that this mass is an
upper limit since at 7~mm we are sensitive to spatial scales of $\sim2500$~AU, smaller than that achieved with the \nh\ observations.
Thus, the temperature should be higher than that estimated from the \nh\ emission, and the resulting mass would be lower. The
uncertainty in the mass is around a factor of 4 mainly due to uncertainties in the dust opacity and the dust emissivity index.


As interferometers are not sensitive to large-scale structures, we compared the 7~mm continuum emission with the 1.2~mm dust emission from
the IRAM\,30\,m telescope. We followed the method described in \citet{girart00}, which relates the FWHM of the single-dish emission with the
half-power $(u,v)$ radius, to estimate the magnitude of this effect at 7~mm. By applying the relation of \citet{girart00}, and adopting a
FWHM for MM1 of $\theta_{\mathrm{FWHM}}\simeq(13''\times10'')^{1/2}=11''$ the half-power $(u,v)$ radius of the 1.2~mm dust emission MM1
becomes $r\simeq8$~k$\lambda$. Given the shortest baseline of the VLA in the D configuration, which is 2.5~k$\lambda<8~k\lambda$, and the
size of the observed emission one can estimate the fraction of correlated flux detected by the interferometer. For a source of 11$''$, this
corresponds to 94~\% for our VLA configuration at 7~mm, indicating that at 7~mm we are not filtering out too much dust emission.




%
%

\subsection{\nh (1,1) and (2,2)}

\begin{figure}[t]
\begin{center}
\begin{tabular}[b]{c}
    \epsfig{file=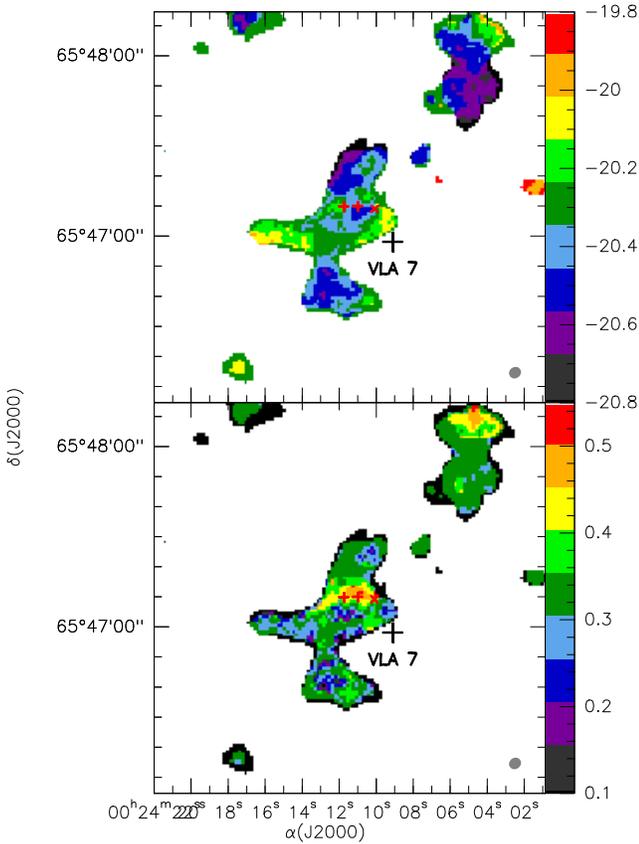,scale=0.75}\\
    \end{tabular}

    \caption{\emph{Top panel:} First-order moment map of the \nh\ $(1,1)$ main line emission. \emph{Bottom panel:} Second-order
moment map of the \nh\ $(1,1)$ main line emission.  Symbols are the same as in Fig.~\ref{fnh3mom0}. Color wedge scales are
\kms. The synthesized beam is shown at the bottom right corner of the images. Note that the second-order moment gives
the velocity dispersion, and must be multiplied by the factor $2\sqrt{2\ln2}\simeq2.35$ to convert to full width to half
maximum.}

\label{fnh3moms}
\end{center}
\end{figure}

\begin{table*}[t]
\caption{\nh\ line parameters obtained from the Gaussian fits to the \nh\,(1,1) and (2,2)}
\begin{center}
\begin{tabular}{lcccccccccc}
\hline\hline
&\multicolumn{2}{c}{Position}
&
&\multicolumn{3}{c}{Main line}
&
&\multicolumn{2}{c}{Inner satellites}
&\\
\cline{2-3}
\cline{5-7}
\cline{9-10}
&$\alpha$ (J2000) &$\delta$ (J2000)
&Transition
&\vel
&$\Tmb$
&$\Delta\,v$
&
&$\Tmb$
&$\Delta\,v$
&$\frac{\Tmb\,(m)}{\Tmb\,(is)}$
\\
Source
&(h m s)
&($\degr$ $'$ $''$)
&$(J,K)$
&(\kms)
&(K)
&(\kms)
&
&(K)
&(\kms)
&
\\
\hline
VLA\,8A &00 24 11.01 &$+$65 47 09.9  &$(1,1)$ &$-19.98\pm0.01$ &3.9$\pm$0.2  &$1.19\pm0.02$ &&1.8$\pm$0.2 &1.14$\pm$0.04  &2.2 \\
	&&			     &$(2,2)$ &$-19.92\pm0.03$ &1.8$\pm$0.1  &$0.83\pm0.07$ &&  &  & \\ 
\hline	
VLA\,8B	&00 24 11.76  &$+$65 47 09.8 &$(1,1)$ &$-19.89\pm0.01$ &3.6$\pm$0.2  &$1.16\pm0.02$ &&1.5$\pm$0.2 &1.18$\pm$0.06 &2.4  \\ 
	&&			     &$(2,2)$ &$-19.88\pm0.04$ &1.5$\pm$0.1  &$0.91\pm0.08$ &&  &  & \\
\hline
IRS\,1	&00 24 10.10  &$+$65 47 09.1 &$(1,1)$ &$-19.97\pm0.02$  &2.2$\pm$0.1  &$1.19\pm0.03$ &&1.0$\pm$0.1 &1.11$\pm$0.07 &2.2\\
	&&			     &$(2,2)$ &$-19.83\pm0.05$  &1.0$\pm$0.1  &$0.92\pm0.08$ &&  &  & \\ 
\hline
MM2	&00 24 12.30  &$+$65 46 45.0 &$(1,1)$ &$-19.84\pm0.02$ &1.4$\pm$0.1  &$1.19\pm0.05$ &&0.9$\pm$0.1 &1.06$\pm$0.05 &1.6 \\ 
	&&  		       	     &$(2,2)$ &$-19.85\pm0.08$ &0.5$\pm$0.1  &$0.88\pm0.17$ &&  &  & \\   
\hline
MM1	&00 24 10.80  &$+$65 47 06.0 &$(1,1)$ &$-19.94\pm0.01$ &4.8$\pm$0.2   &$1.10\pm0.03$ &&2.5$\pm$0.1   &1.02$\pm$0.04   &1.9 \\ 
	&& 			     &$(2,2)$ &$-19.96\pm0.04$ &1.8$\pm$0.1   &$0.83\pm0.10$ &&  &  & \\  
\hline
SC	&00 24 05.30  &$+$65 47 47.0  &$(1,1)$ &$-19.95\pm0.01$ &2.7$\pm$0.1  &$0.97\pm0.02$   &&1.5$\pm$0.1 &0.95$\pm$0.05 &1.8 \\

\hline
\end{tabular}
\end{center}
\label{nh3gauss}
\end{table*}

\begin{table}[t]
\caption{\nh\ line parameters from the fits to the \nh\,(1,1) magnetic hyperfine components}
\begin{center}
\begin{tabular}{lcccc}
\hline\hline
&\vel
&$\Delta\,v^{\mathrm{a}}$
&$A\tau_m^{\mathrm{b}}$
&$\tau_m^{\mathrm{c}}$\\
Source
&(\kms)
&(\kms)
&(K)
&\\
\hline
VLA\,8A  &$-19.99\pm0.01$ &$0.8\pm0.1$ &11.04$\pm$0.32 &2.71$\pm$0.13  \\
VLA\,8B	 &$-19.92\pm0.01$ &$0.9\pm0.1$ &\phn8.41$\pm$0.35 &2.12$\pm$0.16   \\
IRS\,1	 &$-19.92\pm0.01$ &$0.9\pm0.1$ &\phn8.39$\pm$0.34 &2.11$\pm$0.15   \\
MM2	 &$-19.85\pm0.01$ &$0.7\pm0.1$ &\phn6.60$\pm$0.34 &4.40$\pm$0.37 \\
MM1	 &$-19.96\pm0.01$ &$0.7\pm0.1$ &16.34$\pm$0.54 &3.46$\pm$0.17 \\
SC	 &$-19.96\pm0.01$ &$0.6\pm0.1$ &10.08$\pm$0.49 &3.96$\pm$0.27\\

\hline
\end{tabular}
\end{center}
\begin{list}{}{}
\item[$^\mathrm{a}$] Intrinsic line width (FWHM) of the magnetic hyperfine component.
\item[$^\mathrm{b}$] $A=f(J_{\nu}(T_{\mathrm{ex}})-J_{\nu}(T_{\mathrm{bg}}))$, where $f$ is the filling factor, $\Tex$ is the excitation temperature of the
transition, $T_{\mathrm{bg}}$ is the background radiation temperature, and $J_{\nu}(T)$ is the intensity in units of temperature,
$J_{\nu}(T)=(h\nu/k)/(exp(h\nu/kT)-1)$.
\item[$^\mathrm{c}$] Optical depth of the main line obtained from the fit.  
\end{list}
\label{nh3par}
\end{table}

\nh\,$(1,1)$ and \nh\,$(2,2)$ emission is detected in individual velocity channels from $-$19.6 to $-$20.9~\kms\ and from $-$20.2 to
$-$20.9~\kms, respectively. We also detected the inner satellite lines as well as one of the outer satellite lines of the $(1,1)$
transition, and we marginally detected one of the inner satellite lines of the \nh\,(2,2). Figure~\ref{fnh3ch} shows the velocity channel
maps of the \nh\ $(1,1)$ and $(2,2)$ main line emission, and in Fig.~\ref{fnh3spec} we show the \nh\ $(1,1)$ and $(2,2)$ spectra, not
corrected for the primary beam response, at some selected positions. In Table~\ref{nh3gauss} we show the line parameters toward these
positions obtained from a Gaussian fit to the \nh\,(1,1) main and inner satellites lines, as well as the \nh\,(2,2) line. We additionally
show the ratio of the main line to the inner satellites, which gives an indication of the line optical depth. The values obtained for the
optical depth are in the range $\tau_{m}\sim2-3$. In Table~\ref{nh3par} we present the line parameters resulting from the hyperfine fit
to the \nh\,(1,1) line toward the same positions. Given our spectral resolution, $\Delta\,v=$0.6~\kms, the intrinsic line width has been
obtained by selecting the value that minimize the hyperfine fit residual. The optical depths derived from the magnetic hyperfine fit are
systematically higher but compatible with the values obtained from the Gaussian fit (see Appendix for further details). 

\begin{table*}[!ht]
\caption{Summary of \hho\ maser observations toward IRAS 00213+6530}
\begin{center}
\begin{tabular}{lccccccc}
\hline\hline
&Observation
&&&Positional
&&rms
&$I_\mathrm{peak}$
\\
Reference
&date
&Telescope 
&Beam
&accuracy
&Detection
&(\jpb)
&(\jpb)
\\
\hline
Felli \et\ 1992  &1991 Jan 18  &32\,m Medicina  &$1\farcm9$  &$15''$  &no  &1.1  &$<5.5$\\
Anglada \et\ 1997&1990 Feb \phnn\ &37\,m Haystack &$1\farcm4$ &$15''$ & no& 1.6 &$<4.8$\\
Han \et\ 1998    &1993 Nov 21  &13.7\,m Purple Mount &$4.2'$  &$20''$  &yes$^\mathrm{a}$ &9.1 &38.8\\
this work        &2006 Dec 2   &VLA           &$2\farcs1$ &$<1''$  &no  &0.003  &$<0.015$\\
this work        &2008 Apr 18  &NASA\,70\,m   &$40''$  &$10''$  &yes$^\mathrm{b}$ &0.03 &0.111\\
this work        &2008 Jun 19  &NASA\,70\,m   &$40''$  &$10''$  &no &0.04 &$<0.21$\\
this work	 &2008 Sep 23  &NASA\,70\,m   &$40''$  &$10''$  &no &0.06 &$<0.28$\\  	
\hline
\end{tabular}
\end{center}
\begin{list}{}{}
\item[$^\mathrm{a}$] Detection at $-0.7$~\kms.
\item[$^\mathrm{b}$] Marginal detection at $-15$~\kms.
\end{list}
\label{thho}
\end{table*}

\begin{figure}[!ht]
\begin{center}
\begin{tabular}[b]{c}
       \epsfig{file=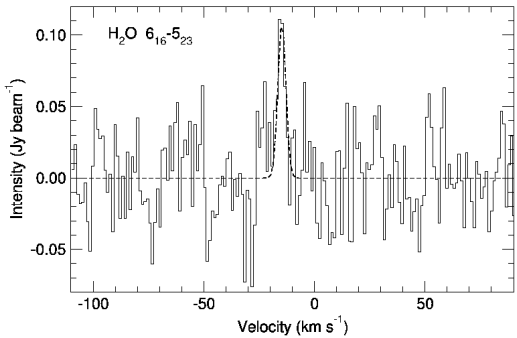,scale=0.35}
       \end{tabular}
        \caption{Spectrum of the \hho\ maser observed with the NASA\,70\,m telescope on 2008
       April 18. The spectrum was smoothed to a spectral resolution of 1.13~\kms. The dashed line is a Gaussian fit to the
       spectrum.}
\label{watermaser}
\end{center}
\end{figure}

In Fig.~\ref{fnh3mom0} we present the zero-order moment map of the \nh\,$(1,1)$ and $(2,2)$ emission (integrated intensity including only the
main line). While the \nh\,$(1,1)$ map shows significant extended structure, the emission of the \nh\,$(2,2)$ is compact around the position
of the centimeter/millimeter sources, We also detected faint \nh\,(2,2) emission toward the northern peak of MM2. The overall structure of
the \nh\,$(1,1)$ emission consists of three clouds. The central \nh\ cloud coincides with the dust condensation MM1. Around $\sim22''$ south
of MM1 there is another cloud detected only in \nh\ (hereafter southern cloud). Moreover, the \nh\ cloud located to the northwest of MM1 is
associated with the dust condensation MM2. Note that, because of the small size of the VLA primary beam at 1.3~cm, the sensitivity toward MM2
is two times lower than toward the center of the field. Then, the dense gas emission traced by the \nh\ molecule roughly follows the 1.2~mm
dust continuum emission, except in the southern cloud in which we did not detect dust emission at all (see Fig.~\ref{fnh3mom0}). While the
\nh\ cloud MM1 engulfs the centimeter source VLA~8 (i.\,e., IRS~1, VLA~8A, and VLA~8B), no \nh\ emission is seen toward the position of
VLA~7, which falls close to the edge of the \nh\ emission. It is important to note that there is a near-infrared source detected in the 2MASS
bands, 2MASS J00241251$+$6546418, spatially coinciding with the southern \nh\ cloud, which probably is not associated with the dense gas (see
Sect.~5.2 for a complete explanation).

In Fig.~\ref{fnh3moms} (top) we show the first-order moment (intensity weighted mean \vel) of the \nh\,$(1,1)$ main line
emission. As can be seen in this figure, the velocity along the central ammonia cloud MM1 shows only small variations, and no
significant velocity gradients are found between MM1 and the southern cloud. The millimeter source VLA~8B is redshifted by
$\sim$~0.3~\kms\ with respect to VLA~8A. In addition toward the western and eastern edges of MM1 the \nh\ emission is
redshifted by $\sim0.4$~\kms. Toward MM2 there is a small velocity gradient in the north-south direction of $\sim0.6$~\kms\
along a region of $\sim30''$. 


The map of the second-order moment (intensity weighted velocity dispersion) of the \nh\,$(1,1)$ main line emission is shown in
Fig.~\ref{fnh3moms} (bottom). The typical value found for the velocity dispersion in the southern cloud is $\sim$\,0.25$-$0.3~\kms, which
corresponds for a Gaussian line profile to a full width at half maximum (FWHM) of 0.6~\kms, similar to the instrumental resolution. In
contrast, in the \nh\ cloud MM1 there is evidence of line broadening toward the three embedded sources (forming an arc-shaped structure),
with values up to 0.4$-$0.5~\kms, corresponding to line widths of 0.7$-$1~\kms, corrected for instrumental resolution, significantly higher
than the expected thermal line width $\sim0.23$~\kms\ (estimated for a kinetic temperature of $\sim20$~K), indicative of a significant
contribution from non-thermal processes. We found that the typical value of velocity dispersion in the \nh\ cloud MM2 is
$\sim$\,0.3$-$0.35~\kms, corresponding to line widths of 0.4$-$0.5~\kms, corrected for instrumental resolution, and rises up to 0.45~\kms\ 
(line width of 0.9~\kms, corrected for instrumental resolution) toward the northern peak of MM2. These values are slightly higher than those
found in the southern cloud, suggesting that the gas is being perturbed in the \nh\ cloud MM2.

Finally, we compared the VLA \nh\,(1,1) emission with the single-dish \nh\,(1,1) emission of \citet{sepulveda2001}. From the VLA
\nh\ observations we estimated a peak intensity of 27.8~m\jpb, which corresponds to a main beam brightness temperature $\Tmb\simeq4$~K. The
largest angular scale detectable by the VLA at 1.3~cm in the D configuration is around $60''$, and the size of the largest features
detected by us with the VLA is $\sim30''$. From the size of the \nh\ emission we can estimate the dilution effect when observed this
emission with a single-dish telescope of $1\farcm4$. The dilution effect would decrease the main beam brightness temperature, $\Tmb$, by a
factor of 7.8, consistent with the $\Tmb$ measured by \citet{sepulveda2001}, indicating that the fraction of emission filtered out by the
interferometer must be small or negligible.



\subsection{\hho\ maser emission}

In Table~\ref{thho} we compiled the \hho\ maser observations carried out toward I00213 up to date, including our observations with the VLA
and NASA\,70\,m. The \hho\ maser emission was clearly detected only in 1993 by \citet{han1998}, using the 13.7~m radio telescope of Purple
Mountain Observatory, with a peak intensity of 38.8~Jy. We did not detect \hho\ maser emission toward I00213, except in the
NASA\,70~m observations on 2008 April 18, where we marginally detected emission at a 3~$\sigma$ level of 0.1~\jpb\ (see
Fig.~\ref{watermaser}). The integrated intensity was $\sim0.49$~\jpb~\kms, and the velocity of the feature was $-$15.2~\kms, offset by
$\sim5$~\kms\ from the velocity of the cloud, $\sim-20$~\kms. 

The characterization of \hho\ maser emission is difficult because of its high temporal variability, and thus it is possible that there was no
maser during the epochs of observation (2006 Dec and 2008) with the exception of the marginal detection during 2008 April. However,
it is not clear whether the \hho\ maser detected by \citet{han1998} is associated with the I00213 region. On one hand, the maser emission was
detected at \vel$=-0.7$~\kms, 19.2~\kms\ offset from the systemic velocity of the cloud studied in this work. On the other hand, the beam of
the 13.7~m radio telescope at this wavelength is 4$\farcm$2, with a pointing accuracy of $20''$, making it difficult to ascertain whether the
maser is associated with the IRAS source. In addition, other attempts to detect \hho\ maser emission toward I00213 have failed
(\citealt{felli1992}; C. Codella observed the source between 1989 and 1999 using the 32m Medicina telescope but did not detect it; private
communication). Thus, the maser activity of I00213 seems to be currently in a rather quiescent state.




\subsection{CCS emission}

\begin{figure}[t]
\begin{center}
\begin{tabular}[b]{cc}
    \epsfig{file=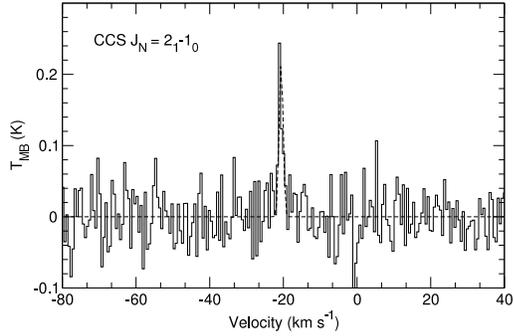,scale=0.35} 
    \end{tabular}
 \caption{Spectrum of the CCS $J_N=2_1$--$1_0$ transition observed with the NASA\,70\,m telescope toward
 I00213. The dashed line is a
 Gaussian fit to the spectrum.}
\label{fccs}
\end{center}
\end{figure}

\begin{table}[t]
\caption{Parameters of the NASA\,70\,m CCS line emission toward I00213\label{tccs}}
\begin{center}
\begin{tabular}{lccc}
\hline\hline
Line 
&$v_\mathrm{LSR}$
&$\Delta v$
&$T_\mathrm{MB}$\\
&(\kms)
&(\kms)
&(K) 
\\  
\hline
CCS~$J_N=2_1$--$1_0$   &$-20.6\pm0.2$  &$1.3\pm0.5$ &0.22$\pm$0.05    \\
\hline
\end{tabular}
\label{tiram30m}
\end{center}
\end{table}

In Table~\ref{tccs} we show the parameters obtained from a Gaussian fit of the CCS line detected with the NASA\,70\,m, after combining the
data of the two days observed, and we show its spectrum in Fig.~\ref{fccs}. The line is centered around $-20$~\kms, the same velocity as \nh,
and its line width (FWHM) is $\sim1.3$~\kms, larger than the expected thermal line width for a kinetic temperature of $\sim20$~K (estimated
from \nh), which is of $\sim0.13$~\kms. This could be indicative of the CCS line width having a strong contribution from non-thermal
processes, such as turbulence injected by the passage of an outflow and/or global systematic motions. The measured line width is higher than
the largest line width measured by \citet{deGregorio06} toward a sample of low-mass YSOs. A high angular resolution study of
\citet{deGregorio05} in CCS shows that this molecule is possibly enhanced via shocked-induced chemistry, and has a velocity gradient in the
same direction of the outflow. 




\section{Analysis}

\subsection{Rotational temperature and column density maps}

\begin{figure}[t]
\begin{center}
\begin{tabular}[b]{c}
	\epsfig{file=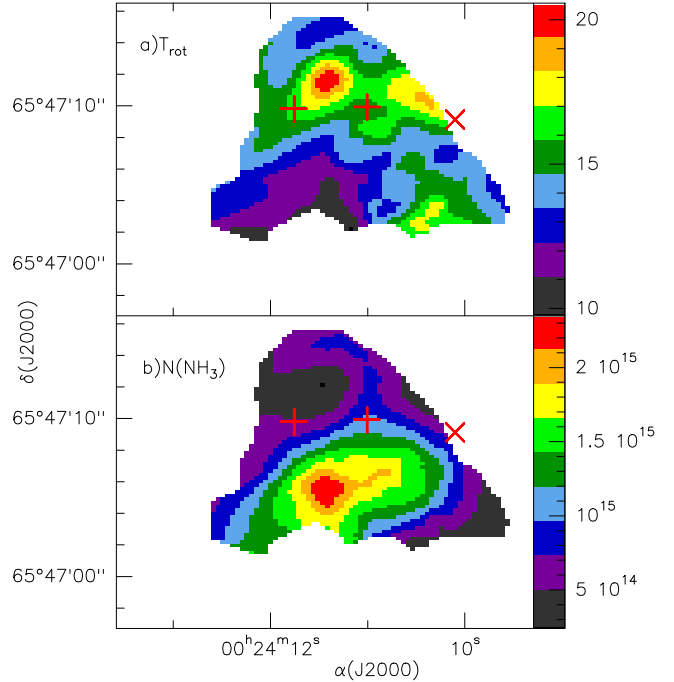,scale=0.8}
       \end{tabular}
   
       \caption{\emph{a)} `Average' Rotational temperature map from \nh\ $(1,1)$ and \nh\ $(2,2)$ toward MM1 (see text). Scale units are in
       K. \emph{b)} \nh\ column density map. Scale units are in \cmd. In both panels red crosses mark the position of the two millimeter
       sources, VLA~8A and VLA~8B, and the red tilted cross marks the position of IRS~1. Note that VLA~7 lies outside the limits of this
       plot.}

\label{fnh3maps}
\end{center}
\end{figure}


We computed maps of the rotational temperature and column density of \nh. To do this, we extracted the \nh\,$(1,1)$ and $(2,2)$ spectra on a
grid of points separated by $1''$ in the \nh\ cloud MM1. We fitted the hyperfine structure of the \nh\,$(1,1)$ and a single Gaussian to the
\nh\,$(2,2)$ for each spectrum. For the \nh\,$(1,1)$ transition we fitted only the spectra with an intensity greater than 5~$\sigma$ in order
to ensure the detection of the satellite lines, whereas for the \nh\,$(2,2)$ we fitted the spectra with an intensity greater than
4~$\sigma$.



From the results of the fits of \nh\,$(1,1)$ and \nh\,$(2,2)$ we computed the rotational temperature ($\Trot$) and \nh\ column density maps
following the standard procedures (\citealt{ho1983,harju1993,sepulveda1993,anglada1995}, see Appendix for a complete description of their
derivation). This analysis assumes implicitly that the physical conditions of the gas are homogeneous along the line-of-sight, i.e., the
excitation and the rotational temperature are constant along the line-of sight. Since gradients are probably present along the line-of-sight
(see Sect.~4.2), the values  obtained from this analysis should be considered as some kind of `average' along the line of sight.




The map of the `average' rotational temperature obtained is shown in Fig.~\ref{fnh3maps}a. Interestingly, to the north of the millimeter
sources VLA~8A and VLA~8B, there is a temperature enhancement, reaching a maximum value of 20~$\pm$~2~K.  Toward the millimeter sources the
`average' rotational temperature is around 16~K, and it decreases toward the south. In addition, at the western edge of the \nh\ cloud MM1
and toward the infrared source IRS~1 there is a temperature enhancement. The `average' rotational temperature obtained at the position of
IRS~1 is $\sim15\pm1$~K, which is consistent with the expected association of this source with the high density gas. We also found a local
maximum of temperature around $8''$ to the southwest of VLA~8A.


Figure~\ref{fnh3maps}b shows the resulting column density map for \nh, obtained after correction for the primary beam response. The highest
values of the \nh\ column density, 2.5$\times10^{15}$~\cmd, are found to the south of the millimeter sources, where the rotational
temperature shows the smallest values. Toward the two millimeter sources, VLA\,8A and VLA\,8B, the \nh\ column density is
$\sim8\times10^{14}$~\cmd.

%
%

\begin{figure}[t]
\begin{center}
\begin{tabular}[b]{c}
     	\epsfig{file=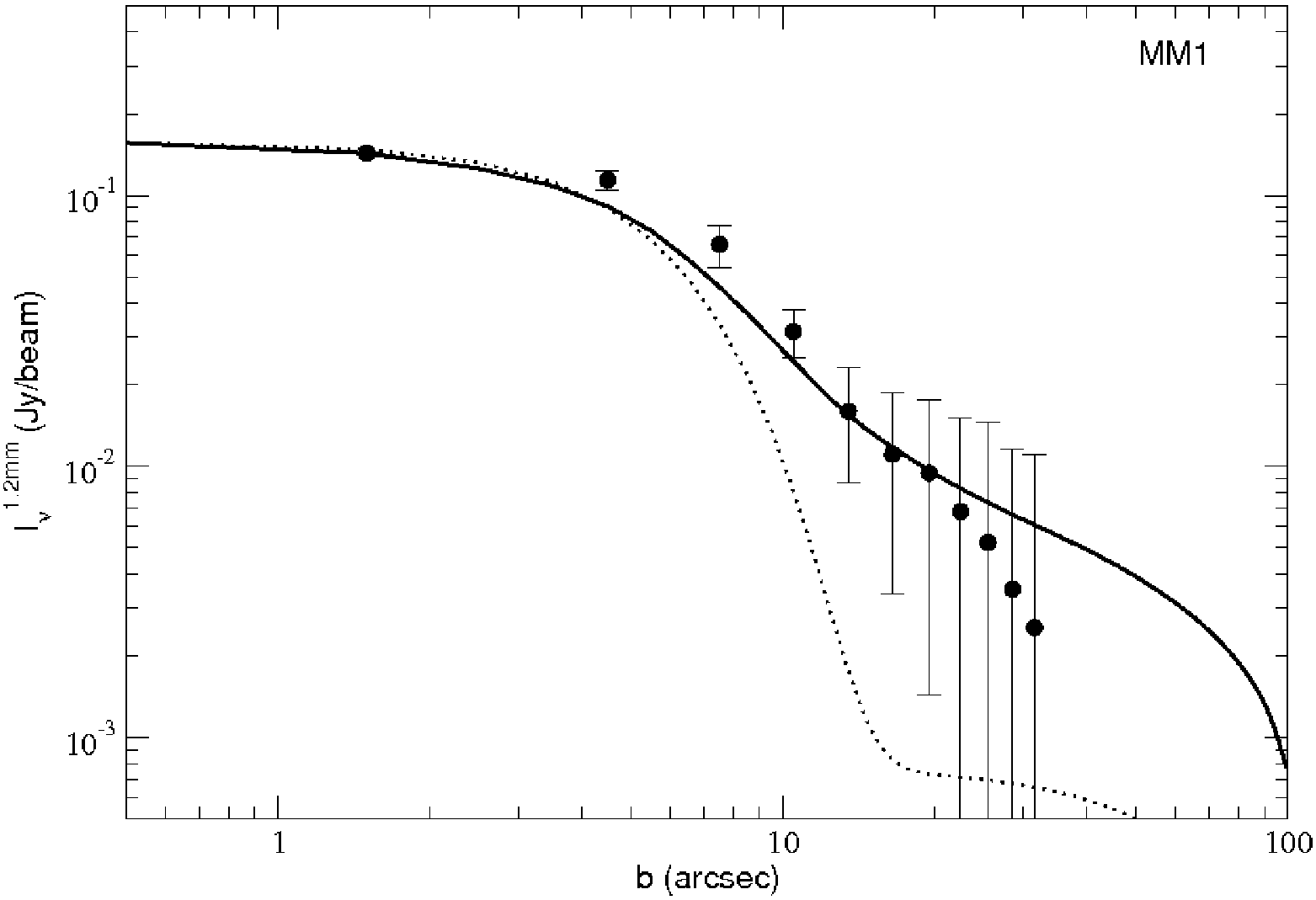,scale=0.32} \\
	 \epsfig{file=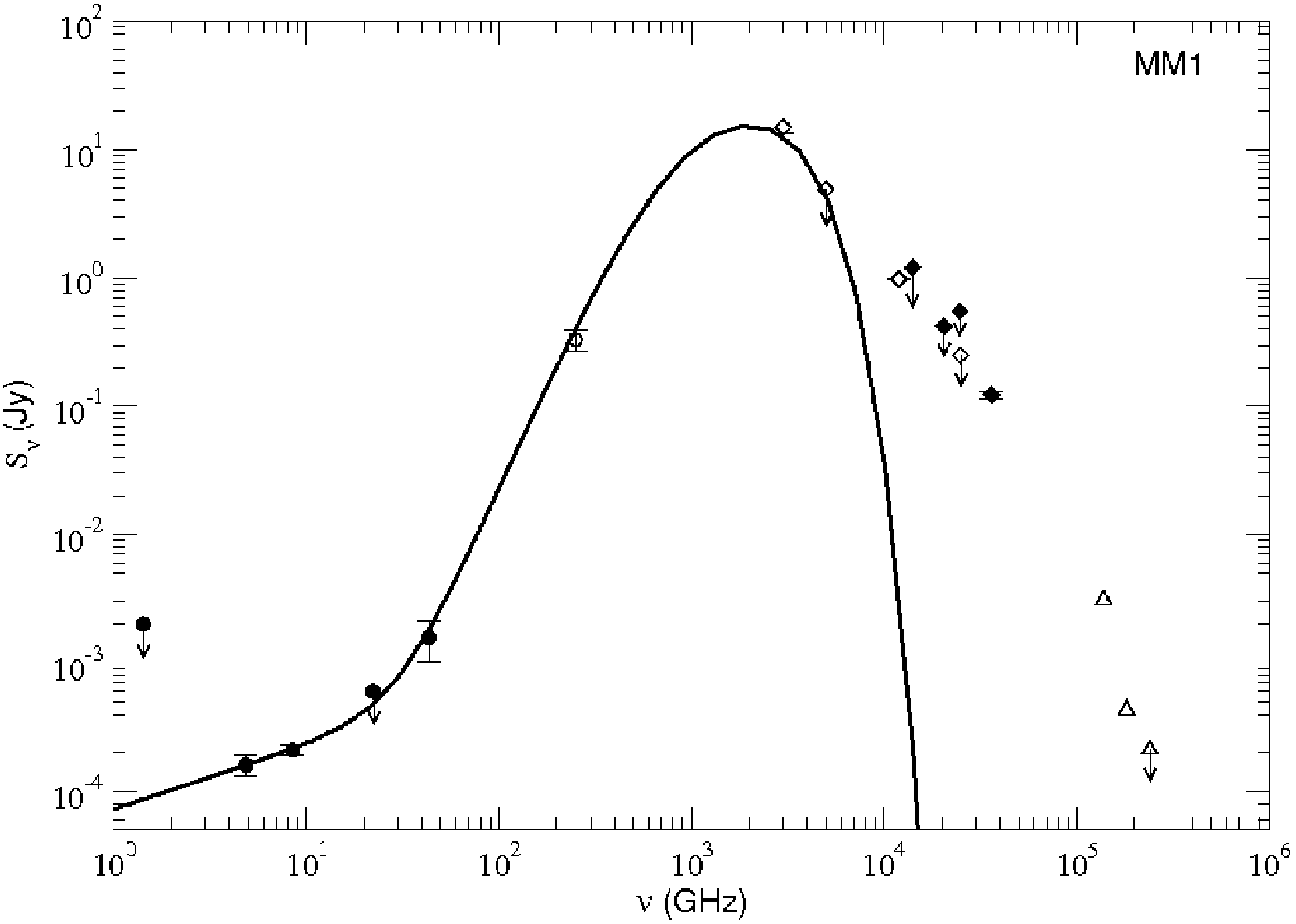,scale=0.32} 
       \end{tabular}

	\caption{\emph{Top:} Circularly  averaged radial intensity profile at 1.2~mm of MM1. The error bars indicate the data rms inside each
       $3''$ wide ring. The dotted line shows the beam profile, including the error beams as given by \citet{greve98}. The solid line shows
       the best fit model. \emph{Bottom:} Spectral Energy Distribution (SED) for MM1 in the IRAS~00213+6530 region. Filled circles are data
       from the VLA, the open circle represents the 1.2~mm (250~GHz) flux from the IRAM\,30\,m telescope, diamonds are IRAS data, filled
       diamonds are MSX data, and triangles are 2MASS data (see Table~\ref{tsed}). The solid line is the sum of the free-free emission and
       envelope dust emission.}

\label{fradprof}
\end{center}
\end{figure}




\subsection{Radial intensity profile and Spectral Energy Distribution}

In order to study the spatial structure of the dusty condensation MM1 detected at 1.2~mm, we computed the circularly averaged radial
intensity profile, in rings of $3''$ width, as a function of the projected distance from the 1.2~mm peak (see
Table~\ref{iram30m}). The result is plotted in Fig.~\ref{fradprof} (top panel), together with the IRAM\,30\,m beam profile, and the
best model fitted to the data. 

In Table~\ref{tmodelprof} we show the parameters of the envelope model, which have been calculated by following \citet{estalella2009}. In
this paper the authors fit the radial intensity profile, using the full Planck function to describe the intensity, and the observed Spectral
Energy Distribution (SED) simultaneously, adopting as a model of the source a spherically symmetric envelope of gas and dust surrounding the
protostar(s). We do not include the possible contribution from the circumstellar disk because our angular resolution, $\sim10''$ (8500~AU),
is much larger than the typical sizes of accretion disks (tens to hundreds of AU). Note that we do not attempt to fit the near/mid-infrared
emission of MM1, since it originates from components at higher temperatures than that responsible for the mm/submm emission. So, the model is
used to fit the SED up to frequencies corresponding to 60~$\mu$m. We assumed the dust opacity law
$\kappa_\mathrm{{\nu}}=0.01(\nu/230~\mathrm{GHz})^{\beta}$ $\mathrm{cm^{2}~g^{-1}}$ \citep{ossenkopf94}, being $\beta$ a free parameter of
the model. For the density and temperature we considered power-laws as a function of radius, $\rho\propto~r^{-p}$ and $T\propto~r^{-q}$, with
$p$ as a free parameter of the model, and $q=2/(4+\beta)$ \citep{Kenyon93}. In order to compare the model with the observed intensity profile
we computed the 2-dimensional intensity map from the model, and we convolved it with the IRAM\,30~m beam. We note that the beam was adopted
to be the sum of two circular Gaussian. Then, from the convolved map we computed the circularly averaged profile. Thus, our free parameters
were the dust emissivity index $\beta$, the  density power-law index $p$, and the scale of the density and temperature power-laws, namely,
the density and temperature at a radius of 1000~AU (taken arbitrarily as the reference radius for the power-laws). From the fitted parameters
we derived the temperature power-law index $q$, the size, $R_{\mathrm{env}}$, and mass, $M_{\mathrm{env}}$, of the envelope.
$R_{\mathrm{env}}$ is defined as the radius for which the envelope density falls to a particle density similar to the ambient density, taken
as $7.3\times10^3$~\cmt. $M_{\mathrm{env}}$ is the integral of the envelope mass density up to the envelope radius $R_{\mathrm{env}}$ (see
Table~\ref{tmodelprof}).

\begin{table}[t]
\caption{Parameters of the envelope model used to fit the radial intensity profile of the 1.2~mm
continuum emission and the SED of MM1}
\begin{tabular}{lc}
\hline\hline
parameter
&value\\
\hline
Dust emissivity index $\beta$  &\phn1.5\phn\\
$T$ power-law index  &\phn0.36$^{\mathrm{a}}$\\
$T$ at $1000$~AU (K)  &31\\
Density power-law index &\phn1.9\phn\\

Density at $1000$~AU (g\,cm$^{-3}$) &1.1$\times10^{-19}$\\
\hline
$R_{\mathrm{env}}^{\mathrm{b}}$ (AU)     &2.1$\times10^{4}$\\
Envelope mass$^{\mathrm{c}}$ (\mo)  &6.0\\
\hline
\end{tabular}
\begin{list}{}{}
\item[$^\mathrm{a}$] $2/(4+\beta)$
\item[$^\mathrm{b}$] Radius for which the envelope density falls to a particle ambient density $n$(\hh) of $7.3\times10^{3}$~\cmt.
\item[$^\mathrm{c}$] Integrated up to $R_{\mathrm{env}}$.
\end{list}
\label{tmodelprof}
\end{table}

\begin{table}[!t]
\caption{Photometry of \textbf{the MM1 clump}}
\begin{center}
{\small
\begin{tabular}{rrrl}
\noalign{\smallskip}
\hline
\hline\noalign{\smallskip}
$\lambda$
&$S_\nu$
&FWHM
&Survey
\\
($\mu$m)
&(mJy)
&($''$)
&Instrument
\\
\noalign{\smallskip}
\hline\noalign{\smallskip}
1.25	&$<0.21^\mathrm{a}$          &2	&2MASS\\
1.65	&0.431$\pm$0.001$^\mathrm{a}$ &2	&2MASS\\
2.17	&3.104$\pm$0.001$^\mathrm{a}$	   &2	&2MASS\\
8.28	&123$\pm$10$^\mathrm{b}$	   &18.3    &MSX\\
12      &$<250$           &50     &IRAS\\
12.13   &$<549$		&18.3	  &MSX\\
14.65   &$<420$		&18.3	  &MSX\\
21.34   &$<1200$        &18.3	  &MSX\\
25	&980$\pm$90	   &50	&IRAS\\
60	&$<4910$          &100     &IRAS\\
100     &15000$\pm$1500      &230     &IRAS\\
1200	&640$\pm$70        &11      &IRAM\,30\,m\\
7000    &1.6$\pm$0.5$^{\mathrm{c}}$       &3       &VLA\\
13000   &$<$0.6$^\mathrm{c, d}$          &1.5     &VLA\\
36000   &0.27$\pm$0.02$^\mathrm{e}$     &12      &VLA\\
60000   &0.13$\pm$0.02$^\mathrm{e}$     &15      &VLA\\
210000   &$<2.0^\mathrm{d}$           &45      &VLA$^\mathrm{f}$  \\
\hline
\end{tabular}
\begin{list}{}{}
\item[$^\mathrm{a}$] J00241110+6547095. 
\item[$^\mathrm{b}$] G120.1368+03.0617.
\item[$^\mathrm{c}$] VLA~8A.
\item[$^\mathrm{d}$] 4$\sigma$ upper limit.
\item[$^\mathrm{e}$] VLA~8 (VLA~8A and VLA~8B).
\item[$^\mathrm{f}$] From the NRAO VLA Sky Survey \citep{condon98}.
\end{list}
}
\end{center}
\label{tsed}
\end{table}

As mentioned above, the model fits the radial intensity profile and the SED simultaneously. The SED for MM1 was built by
using the data shown in this work and by searching the literature for 2MASS, IRAC-Spitzer, MSX, and IRAS data. In Table~\ref{tsed}
we list the photometry used for MM1 (adopting the values of VLA~8A for 1.3~cm, 7~mm and  2MASS, where the high angular resolution
allows us to disentangle the different sources). We were able to fit simultaneously the SED and the intensity profile at 1.2~mm. In
Fig.~\ref{fradprof} (bottom panel) we show the best fit of the model to the observed SED. In this figure we show the sum of the
free-free emission and the envelope flux density integrated inside the radius of the envelope. The centimeter continuum emission is
dominated by free-free emission with a spectral index  of $0.5\pm0.4$. At millimeter and submillimeter wavelengths the dust emission
of the envelope is dominant.



The model resulting from the simultaneous fit to the radial intensity profile and the SED is able to fit remarkably well the radial intensity
profile, but underestimates slightly the intensity at  projected distances of $b=4\farcs5$ and $7\farcs5$ (see Fig~\ref{fradprof} top panel).
This can be due to the presence of several YSOs close ($<10''$) to the 1.2~mm peak. The value obtained for the power-law density distribution
index $p=1.9$ is similar to that found for other protostellar envelopes. \citet{chandler2000} carried out a submillimeter survey of Class~0
and Class~I sources and fitted the observed radial intensity profiles with density index $p$ between 1.5 and 2 for the majority of the
sources. \citet{Hogerheijde00} find $p=0.9$--2.1 in a sample of four Class~I YSOs, and \citet{motte2001} find $p=1.2$--2.6 in a sample of
embedded YSOs in Taurus and Perseus, similar to the value obtained here.

\begin{table}[t]
\caption{Physical parameters of IRAS\,00213$+$6530
}
\begin{center}
\begin{tabular}{lcccc}
\hline\hline
& Mass$^{\mathrm{a}}$ &$\Delta v^{\mathrm{b}}$ &$\theta\,^{\mathrm{c}}$
&$M_{\mathrm{vir}}^{\mathrm{d}}$ \\
Region  &  (\mo)  &(\kms) & (arcsec) &(\mo)\\	
\hline
MM1     &\phb3.5   &0.85  &\phn7&  1.3\\
Southern cloud        &< 0.3         &0.65  &\phn6&  0.7\\ 
MM2   &\phb1.1  &0.75  &11&  1.6\\
\hline
\end{tabular}
\begin{list}{}{}

\item[$^\mathrm{a}$] Estimated as $M=\mu_{\mathrm{H_2}}m_{\mathrm{H}}\sum N(\mathrm{H_{2}})\Delta A$, where $N(H_{2})$ is taken from the \hh\
column density map, $\Delta A$ is the pixel area, and a mean molecular mass per \hh\ molecule  $\mu_{\mathrm{H_2}}=2.8$, which corresponds to
a 10~\% helium abundance. 
\item[$^\mathrm{b}$] Intrinsic line width of the magnetic hyperfine components. 
\item[$^\mathrm{c}$] Deconvolved geometrical mean size of the major and minor axes of the source, obtained from a 2D Gaussian fit.
\item[$^\mathrm{d}$] Estimated for a density distribution $n\propto$~$r^{-2}$.
\end{list}
\label{mass}
\end{center}
\end{table}

\subsection{Column density maps and mass}

\begin{figure*}[!ht]
\begin{center}
\begin{tabular}[b]{cc}
       \epsfig{file=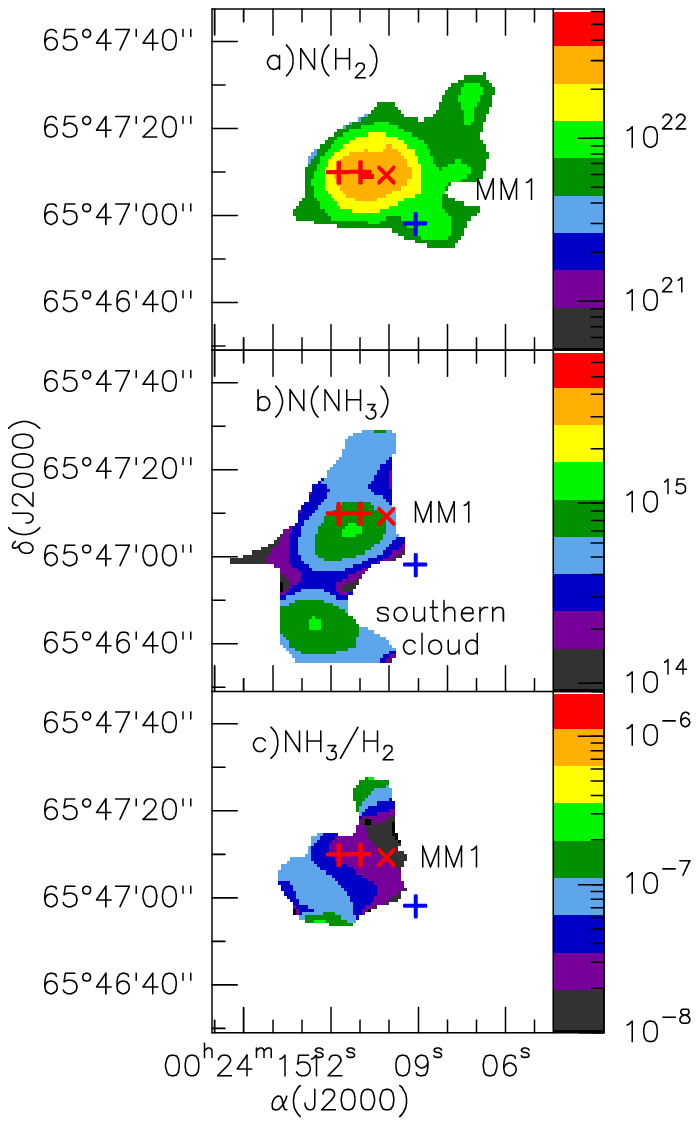,scale=1.25} &
	\epsfig{file=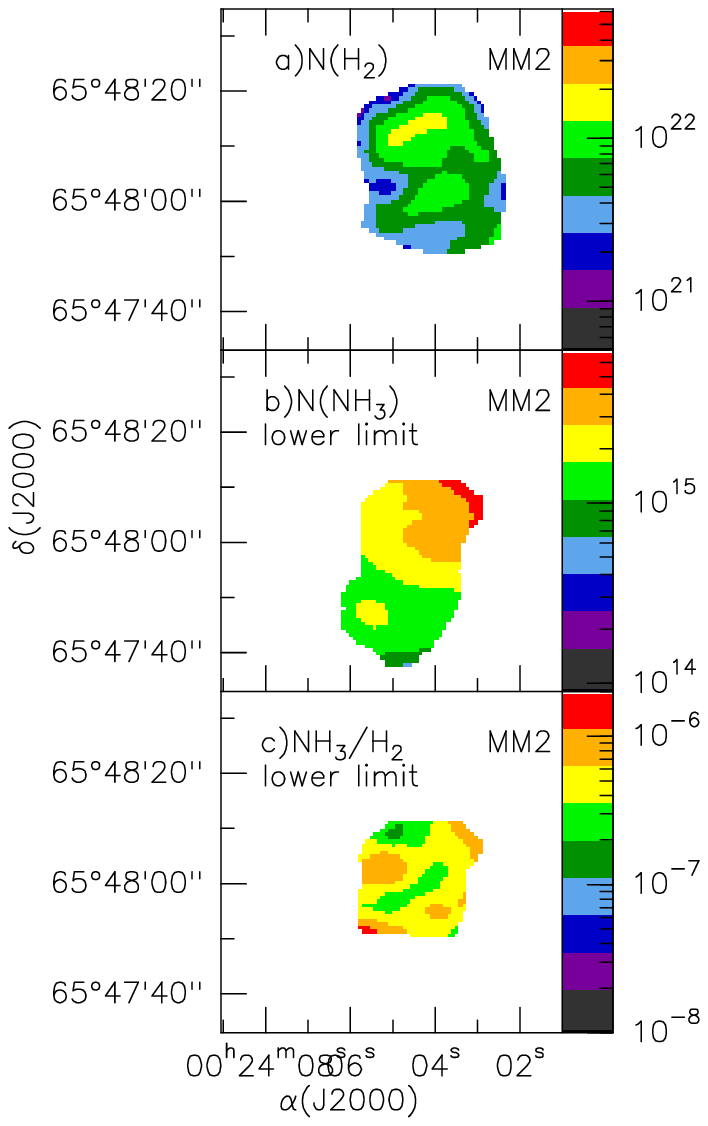,scale=1.25} \\
       \end{tabular}
      \caption{\emph{a)} \hh\ column density map from the 1.2~mm dust emission. \emph{b)} \nh\ column density map.
 \emph{c)} \nh/\hh abundance map, for MM1 and the southern cloud (left panels) and for MM2 (right
 panels) in logarithmic scale. Scale units for the \hh\ and \nh\ column density are \cmd. In all panels red crosses indicate the position
 of the two millimeter sources, VLA\,8A and VLA\,8B, and the red tilted cross marks the position of the infrared
 source IRS~1. The blue cross denotes the position of VLA~7. The color scale is the same for the left and right panels.}
\label{nh3abundance}
\end{center}
\end{figure*}

In order to estimate the relative \nh\  abundance in I00213, we computed the \hh\ column density map. To do so, we used the expression given
in \citet{motte98} to compute the column density of \hh\ from the 1.2~mm dust in a grid of $3\farcs3\times3\farcs3$ (the pixel size of our
maps), and using a dust mass opacity coefficient $\kappa_{\mathrm{1.2mm}}=0.01$~cm$^{2}$\,g$^{-1}$. Since the beam size of the 1.2~mm
emission is a factor 2 larger than the beam size of the \nh\ observations, we convolved the \nh\ emission to the same angular resolution as
the 1.2~mm dust emission ($\sim10''$). We obtained the rotational temperature and column density of \nh\  following the procedure described
in Sect.~4.1 (see Appendix). In MM1 we obtained the rotational temperature from the \nh\,(1,1) and (2,2) emission, which was
converted to kinetic temperature using the relation given in \citet{tafalla2004} (see also Appendix). In order to estimate lower limits of
the \nh\ column density in the southern cloud and MM2 we adopted a rotational temperature of 10~K (see Sect.~5.2).  

Maps of the \hh\ column density and \nh\ column density are shown in Fig.~\ref{nh3abundance}. The \hh\ column density has a maximum
value of $\sim4\times10^{22}$~\cmd\ toward the peak position of the 1.2~mm emission of MM1, and decreases in the more extended structure
to values of $\sim10^{22}$~\cmd. The uncertainty in the \hh\ column density is around a factor of 2, mainly due to the
uncertainty in the dust opacity, as the uncertainty in the mm flux density (the rms of the map), and in the dust temperature constitute
only a small contribution to the total uncertainty. The \nh\ column density map, corrected for the primary beam response, also
shows differences between the three \nh\ clouds (see Fig~\ref{nh3abundance}b). In MM1, the maximum value of the \nh\ column density,
$\sim1.5\times10^{15}$~\cmd, is found to the south of VLA\,8A and VLA\,8B, which is consistent with the results found in Sect.~4.1.
The uncertainty in the \nh\ column density is of the order of $10-20$~\%. It has been  estimated from the uncertainty in
rotational temperature, in line width, and in $A\tau_m$ (see Appendix). The values of the \nh\ column density toward MM2 are
significantly higher, in the range of $(2-7)\times10^{15}$. For the case of MM2 and the southern cloud, the \nh\ column density derived
is a lower limit since the rotational temperature adopted is an upper limit due to the non-detection of \nh\,(2,2).



We estimated the mass of each condensation from the \hh\ column density maps (see Table~\ref{mass}). The mass in MM1 is $\sim$3.5~\mo,
consistent with the values obtained from the 7~mm continuum emission and for the envelope model ($\sim6$~\mo). In addition, we compared the
values obtained with  the virial mass, \mvir, estimated using Eq.~(5) of \citet{beltran06}, which assumes a spherical cloud with a power-law
density distribution $\rho\propto\,r^{-p}$, with p$=2.0$, and neglecting contributions from magnetic fields and surface pressure. As can be
seen in Table~\ref{mass}, the mass of MM1 is higher than the virial mass indicating that it might be unstable and undergoing collapse. The
mass of MM2 is $M\simeq0.7$~\mvir, indicating that the material in this cloud is stable. In contrast, toward the southern cloud the total
mass of gas is $M<0.4$~\mvir, being this clump gravitationally unbound and could disperse at roughly the internal sound speed of
$\sim0.3$~\kms\ on a timescale of around $10^5$~yr, unless it is confined by external pressure. All this suggests that it could be in an
earlier evolutionary stage than MM1 and MM2.


\subsection{Relative \nh\ abundance}

In Fig.~\ref{nh3abundance}c we show the relative \nh\ abundance maps of MM1 and MM2. The typical value of the \nh\ abundance found in MM1, in
a position just south of the three YSOs, is around $2\times10^{-8}-~4\times10^{-8}$, which is similar to the typical value found in
dense clouds (\citealt{herbst1973}, see Anglada \et\ 1995 for a discussion on \nh\ abundances). In MM1, near the YSOs, there is a slight
trend in the \nh\ abundance to increase, from $1\times10^{-8}$ near IRS~1, up to $3\times10^{-8}$ toward  VLA~8B. In this case, the
most evolved YSO IRS~1 is likely dispersing the dense gas material of its surroundings, and the result is a decrease of the \nh\ abundance.
It is, however, a modest effect, slightly above the typical uncertainty, of a factor of 2. Toward the southern cloud we derived a
lower limit for the relative \nh\ abundance of $\sim(1-4)\times10^{-7}$.

Regarding the \nh\ abundance in MM2, we find values ranging from $1-2\times10^{-7}$ up to $1\times10^{-6}$. This abundance is higher than in
MM1, despite the observed \nh\ emission in MM2 being fainter than in MM1, mainly due to two effects: the \nh\ column density is higher due to
the correction for the primary beam response, and the \hh\ column density is low, $\sim3\times10^{21}-1\times10^{22}$~\cmd. It is important
to remark that the relative \nh\ abundance in MM2 is a lower limit. Although these abundances are high, \citet{benson1983} and
\citet{ohishi92} find \nh\ abundances around $\sim2\times10^{-7}$ in starless cores of the dark clouds L1498 and L1512, and in L134N,
respectively. In addition, chemical models of \citet{Hartquist2001} are able to reproduce such a high \nh\ abundance for a young core, Core D
in TMC-1, with N$_2$ and CO not freezing-out. Therefore, it is plausible that MM2 and the southern cloud are in fact evolutionary young, as
suggested by the starless properties of this cloud (see Sect.~5.2).




\section{Discussion}

The results obtained for IRAS~00213+6530 show that in this region there are different sources, which have different radio and
infrared properties:

\subsection{VLA~8A, VLA~8B, and IRS\,1: Multiple sources in different evolutionary stages}

The centimeter and millimeter continuum observations, together with near-infrared data, allowed us to identify three YSOs in MM1: IRS~1,
VLA~8A and VLA~8B. While at 3.6~cm we could not resolve the centimeter emission of VLA~8, at 7~mm we found two sources VLA~8A and VLA~8B. In
the near infrared, only VLA~8A and IRS~1 are detected. In addition, all sources are deeply embedded in the dense gas traced by \nh. Line
broadening and local heating have been detected toward their position, indicating the true association of the three objects with the molecular
gas. 

From these preliminary results we can make a rough estimate of the evolutionary stage of the detected sources. First, IRS~1 is associated
with an infrared source with no detected 7~mm continuum emission. VLA~8A, although bright in the infrared, is the strongest millimeter source
in the field. Its SED has a quite steep profile at the 2MASS wavelengths. Thus, d(log$\lambda$F$_\lambda$)/d(log$\lambda)>0$ between 1 and
10~$\mu$m, which is consistent with the classification of VLA~8A as a Class~I source (\eg\ \citealt{hartmann1998}).  Regarding the
near-infrared colors of the 2MASS sources associated with IRS~1 and VLA~8A, derived from 2MASS photometry (see Table~\ref{2masscolors}), we
found that both sources fall inside the area corresponding to YSOs of Class~0/I (\eg\ \citealt{itoh1996,ojha2004,matsuyanagi2006}).  Finally,
VLA~8B, which is also associated with dense gas tracers, shows no infrared emission at all. At 7~mm VLA~8B has little dust continuum emission
associated, suggesting that the dust emission is colder, probably detectable at 1~mm and/or submillimeter wavelengths. Thus, VLA~8B
could be in a previous stage of evolution, being still more embedded than VLA~8A, possibly in the Class~0 phase. It is worth noting
that we found an extended temperature enhancement to the north of VLA~8A and VLA~8B, which seems to be associated with the passage of an
outflow that heats and perturbs the dense gas: Busquet \et\ (in prep.) find that the large scale molecular outflow \citep{yang1990}, when
observed with high angular resolution, is centered on a position near VLA~8A and VLA~8B, being these sources the candidates to be driving the
observed high-velocity gas.


Thus, it seems that the I00213 region is harboring a multiple system of low mass protostars, indicating that the star formation process in
this region is not producing a single YSO.  Since low mass protostars evolve approximately at the same rate to the main sequence, the
different evolutionary stages found in I00213 suggest that stars in this region are not forming simultaneously but continuously. Actually,
there may be different generations due to different timescales of core collapse, as has been found in other low mass star-forming regions
(\eg\ L1551: \citealt{moriarty2006}), indicating that the formation of different stars is not simultaneous but sequential in time,
maybe triggered by the interaction of the molecular outflow with a dense core in its surroundings \citep{yokogawa03,shimajiri2008}. Therefore,
the initial assumption that star formation occurs in an isolated mode may not be appropriate to describe the I00213 region when the
region is studied with high angular resolution. This poses the question of what extent can we adopt the isolated mode in the theories of star
formation, as this high angular resolution study together with a large number of recent studies (\eg\
\citealt{huard1999,djupvik2006,teixeira2007,carrasco-gonzalez2008,chen2008a,Chen2009,forbrich2008,gutermuth2008,swift2008,girart2009}) suggest
that isolated star formation seems to be rare in the Galaxy, even in low mass star-forming regions.

It is interesting to note that the three YSOs are spatially ordered from youngest (east) to oldest (west), suggesting that an external agent
could be inducing star formation in MM1. For this, it would be very useful to identify and map the molecular outflows in the region.
Alternatively, it would be very useful as well to study the possible association of I00213 with the HII region S171, located to the northwest
of I00213. Finally, it is worth noting that the I00213 region falls exactly on the southern border of the Cep~OB4 shell \citep{kun2008}.

\begin{table}[t]
\caption{Infrared excess of 2MASS sources}
\begin{center}
{\small
\begin{tabular}{lccc}
\noalign{\smallskip}
\hline\hline\noalign{\smallskip}
2MASS
&Associated
&&\\
Source
&Source
&$(J-H)$
&$(H-K)$
\\
\noalign{\smallskip}
\hline\noalign{\smallskip}
J00241110$+$6547095\, &VLA~8A & 1.02$\pm$0.16 &1.47$\pm$0.18 \\
J00241010$+$6547091\, &IRS~1 & 1.25$\pm$0.26 &2.61$\pm$0.27 \\
J00241251$+$6546418   &\ldots	   & 0.48$\pm$0.04 &0.12$\pm$0.05 \\ 
\hline
\end{tabular}
}
\end{center}
\label{2masscolors}
\end{table}

\subsection{Starless candidates: MM2 and the southern cloud}

While the southern cloud is detected only in \nh, MM2 is detected both in \nh\ and dust emission. Given that the $(2,2)$ line is not detected
toward southern cloud, and only detected toward the northern peak of MM2, these clouds are cold, since $T_{\mathrm{rot}}<$11~K. As mentioned
in Sect.~3.3, the near-infrared source  2MASS J00241251$+$6546418 spatially coinciding with the southern cloud is not likely associated with
the dense gas. We estimated the infrared excess from the $(J-H)$ vs $(H-J)$ diagram (see Table~\ref{2masscolors}). The near-infrared colors
derived are characteristic of main sequence stars, giants, supergiants, Class~III sources, or Class~II sources with small infrared excess
(\citealt{itoh1996,matsuyanagi2006}). In addition, this source has an optical counterpart seen in the DSS2 image. Thus, it is likely a
foreground source, not associated with the \nh\ dense gas and the I00213 star-forming region. Regarding MM2, no near-infrared sources are
associated with it. In addition, neither MM2 nor the southern cloud seem to be associated with molecular outflows (Busquet \et\ in prep.).
Therefore, no clear signposts of stellar activity are found for these two clouds, suggesting they could be starless.


\subsection{On the nature of VLA~7}

VLA~7, lying outside the \nh\ dense gas and dust emission, has a very negative spectral index at centimeter wavelengths
($-1.6\pm0.2$), indicating that the emission has a non-thermal origin, found typically for extragalactic sources with steep
spectrum and some pulsars (\eg\ \citealt{lehtinen2003}). Given the close proximity to MM1, we considered the
possibility that VLA~7 could be related to the I00213 region. In low mass star-forming regions non-thermal emission has been
detected toward some YSOs, like Class~0 YSOs \citep{choi08} or T-Tauri stars (\eg\
\citealt{andre1996,rodriguez1999,gibb1999}). However, centimeter emission arising from weak T-Tauri stars is usually
polarized, and they are often optically visible. We did not find evidences of circular polarized emission toward
VLA~7, neither a visible counterpart, so we consider unlikely this possibility. Another possibility to explain the
negative spectral index of VLA~7 is non-thermal synchrotron emission produced in shocked regions of outflowing gas,
found mainly in high mass star-forming regions (\eg\ \citealt{rodriguez89,garay03,rodriguez05}). As there is molecular
outflow emission in this region (Busquet \et\ in prep.) we can not rule out this possibility, and further observations would
help to confirm the truly association of VLA~7 with the I00213 star-forming region.

\section{Conclusions}

We observed with the VLA, IRAM\,30~m Telescope, and the NASA\,70~m antenna the continuum emission at 6~cm, 3.6~cm, 1.3~cm, 7~mm, and 1.2~mm,
the \nh\,(1,1) and \nh\,(2,2) lines, and the \hho\ maser and CCS emission toward the low mass star-forming region IRAS~00213$+$6530. Our main
conclusions can be summarized as follows:

\begin{enumerate}

\item The 1.2~mm continuum emission observed with the IRAM\,30~m shows two dust condensations, MM1 and MM2. The continuum emission at
centimeter and millimeter wavelengths, together with the available data from 2MASS, have revealed three sources, IRS~1, VLA~8A, and VLA~8B,
all embedded in the dusty cloud MM1. These sources show different radio and infrared properties, and seem to be in different evolutionary
stages, with VLA~8B being in the earliest phase. In MM1, low mass star formation appears to proceed along a west-east direction.


\item We marginally detected \hho\ maser emission toward I00213 with the NASA\,70\,m antenna during the observations carried out on 2008
September 23, but other attempts, including the VLA observations, yielded negative results.

\item The YSOs found in the region are deeply embedded in the high-density gas. The \nh\,(1,1) emission traces an elongated structure that
consists of a main cloud (MM1 and the southern cloud) and a smaller cloud, MM2, located to the northwest of MM1.  While the southern cloud
and MM2 appear as quiescent and starless, in MM1 there is evidence of a perturbation of the gas (line broadening and local heating) along the
east-west direction, associated with IRS~1 and north of the two millimeter sources, elongated in the north-south direction. We propose that
part of the dense gas is being perturbed by the passage of one or more outflow(s).


\item We detected CCS emission toward I00213 using the NASA\,70\,m antenna with a line width, $\sim1.3$~\kms, large compared with
previous studies.

\item The source VLA~7, which has a negative spectral index, lies outside but near the border of the \nh\,(1,1) condensation. Although this
source could be a background source we can not rule out the possibility that VLA~7 could be in fact the result of the interaction of a
molecular outflow with the surrounding medium. 

\item We used a spherically symmetric envelope model that fits simultaneously the observed SED from 7~mm to 60~$\mu$m, and the radial
intensity profile at 1.2~mm of the clump associated with MM1. The best fit was obtained for a dust opacity law index $\beta=1.5$, a
temperature at 1000~AU of 31~K, and a density at 1000~AU of $1.1\times10^{-19}$~g~\cmt\ or a particle density of
$n$(\hh)$\sim2.3\times10^4$~\cmt. The envelope radius is $R_{\mathrm{env}}\sim21000$~AU (3.1$\times10^{17}$~cm), and inside this radius the
envelope model mass is 6~\mo.



\item There is a strong differentiation of \nh\ abundance in the region. In particular, we found low values,
$\sim2\times10^{-8}$ of the \nh\ abundance associated with MM1, which contains the YSOs. On the other hand, toward those clouds
with starless properties (the southern cloud and MM2) the \nh\ abundance rises up to $\sim1\times10^{-6}$, suggesting
that in evolved clouds with star-formation activity there is a decrease in the \nh\ abundance.


\end{enumerate}

\begin{acknowledgements}

G.B. is grateful to Serena Viti, David A. Williams, and Oscar Morata  for useful discussion on the ammonia abundance. We
are grateful to an anonymous referee and to the editor for valuable comments. The authors are supported by the Spanish MEC grant
AYA2005-08523-C03, and the MICINN grant AYA2008-06189-C03 (co-funded with FEDER funds). A.P. is also supported by
the MICINN grant ESP2007-65475-C02-02 and the program ASTRID S0505/ESP-0361 from La Comunidad de Madrid and the European Social
Fund. G.A. acknowledges support from Junta de Andaluc\'{\i}a. This publication makes use of the data products from the Two Micron
All Sky Survey, which is a joint project of the University of Massachusetts and the Infrared Processing and Analysis
Center/California Institute of Technology, funded by the National Aeronautics and Space Administration (NASA) and the National
Science Foundation.

\end{acknowledgements}

\appendix

\section{Derivation of $\Trot$ and $N$(\nh) from \nh(1,1) and (2,2) observations}

\subsection*{$T_{\mathrm{MB}}\,(1,1;m)$ and $N\,(1,1)$:}

The \nh\,(1,1) method of CLASS fits the magnetic hyperfine structure of \nh\,(1,1). The output parameters for the fit to the
hyperfine structure are: $A\tau_\mathrm{m}$, the velocity of the reference line, the intrinsic line width, and the optical depth of the (1,1)
main line, $\tau_\mathrm{m}$ (sum of the optical depths of the magnetic hyperfine components of the main line,
$\tau_\mathrm{m}=\tau\,(1,1)/2$. The parameter $A$ is defined as, according to \citet{pauls1983}, $A=f[J_\nu(\Tex)-J_\nu(T_\mathrm{bg})]$,
where $f$ is the filling factor. Then, from the output parameters, and applying the radiative transfer equation, one can obtain the (1,1)
main line temperature $T_{\mathrm{MB}}\,(1,1;m)$,

\begin{equation}
T_\mathrm{MB}\,(1,1;m)=
A\tau_\mathrm{m}\frac{1-e^{-\tau_\mathrm{m}}}{\tau_\mathrm{m}}.
\label{eTm1}
\end{equation}

The excitation temperature $\Tex$ is \emph{not} directly obtained from the fit, but is isolated from the output parameter
$A\tau_\mathrm{m}$,

\begin{equation}
\Tex=\frac{1.14}{\ln\big(1+{1.14}/
{[A\tau_\mathrm{m}/\tau_\mathrm{m}+J_\nu(T_\mathrm{bg})]}\big)}.
\label{eTex}
\end{equation}
Note that no assumption is made concerning $\Tex$ with respect to the background temperature $T_\mathrm{bg}$.

The beam averaged column density in the (1,1) level \citep{anglada1995},

\begin{equation}
\left[\frac{N(1,1)}{\mathrm{cm^{-2}}}\right]=1.58 \times 10^{13}
f
\frac{e^{1.14/\Tex}+1}{e^{1.14/\Tex}-1}
\tau_\mathrm{m}
\left[\frac{\Delta v}{\mathrm{km~s^{-1}}}\right],
\label{en11}
\end{equation}
the filling factor $f$ being assumed to be 1 for our VLA observations.

To derive Eq.~(\ref{en11}), $N(1,1)$ is not approximated to $2N_+(1,1)$, but is taken as
$N(1,1)=N_+(1,1)[1+\mathrm{exp}(h\nu_{11}/k\Tex)]$ (see Harju \et\ (1993) for more details).

\subsection*{$T_{\mathrm{MB}}\,(2,2;m)$:}

For \nh\,(2,2) we fitted one single Gaussian, being the (2,2) main line temperature, $T_\mathrm{MB}\,(2,2;m)$, an output
parameter of the fit.

\subsection*{$T_{\mathrm{rot}}^{21}$:}

The rotational temperature derived from \nh\,(1,1) and \nh\,(2,2) can be estimated, following Ho \& Townes (1983,
Eq.~4), by assuming that the transitions between the metastable inversion doublets are approximated as a two-level
system, and that the excitation temperature $\Tex$ and line width $\Delta v$ are the same for both \nh\,(1,1) and
\nh\,(2,2). Then,

\begin{equation}
T_\mathrm{rot}^{21}=\frac{-41.5}
{\ln\Big(-\frac{0.283}{\tau_\mathrm{m}}\,
\ln\left[1-\frac{T_\mathrm{MB}\,(2,2;m)}{T_\mathrm{MB}\,(1,1;m)}.
(1-e^{-\tau_\mathrm{m}})\right]
\Big)},
\label{etrot}
\end{equation}
Note that we did \emph{not} assume that the emission is optically thin. The assumption of a two-level system is reasonable because
transitions between the metastable inversion doublets are usually much faster than those to other rotational states \citep{ho1983}. If the
density and temperature were high enough to populate the upper non metastable states, multilevel statistical calculations would be required
(\eg\ \citealt{Sweitzer1978}).

An estimate of the gas kinetic temperature can be obtained by correcting the rotational temperature derived from \nh, using the
expression given in Tafalla \et\ (2004),

\begin{equation}
T_\mathrm{k}=
\frac{T_\mathrm{rot}^{21}}{1-\frac{T_\mathrm{rot}^{21}}{42}
\ln\left[1+1.1e^{-16/T_\mathrm{rot}^{21}}\right]},
\label{eTk}
\end{equation}
which is almost independent of core density and size. This relation is recommended for the range
$T_\mathrm{k}=5$--20~K.

\subsection*{$N$(\nh):}

The \nh\ column density was derived by following \citet{ungerechts86}, and Harju \et\ (1993). The main assumptions are:  \emph{i)} only
metastable levels are populated;  \emph{ii)} $\Trot$ is the same for each pair of rotational levels; \emph{iii)} the ratio of the column
densities of each rotational level is the same to the ratio of the column densities of upper inversion levels; \emph{iv)} the
contribution to the total \nh\ column density comes essentially from levels with $J \le 3$; \emph{v)} the relative population of all
metastable levels of both ortho and para-\nh\ is that given by thermal equilibrium at temperature $\Trot$; and \emph{vi)} the frequencies
for the \nh\,(1,1) and \nh\,(2,2) transitions are very similar. With these assumptions,


\begin{equation}
N(\mathrm{NH_3})=N(1,1)\left[
\frac{1}{3}e^{23.4/T_\mathrm{rot}^{21}}
+1
+\frac{5}{3}e^{-41.5/T_\mathrm{rot}^{21}}
+\frac{14}{3}e^{-101.2/T_\mathrm{rot}^{21}}
\right],
\label{ecoldens}
\end{equation}

\subsection*{Uncertainty in $T_{\mathrm{rot}}^{21}$:}

In order to estimate the uncertainties associated with $\Trot$ and $N$(\nh) introduced by this method, we did the
following.

The error of $\Trot^{21}$ was estimated by assuming optically thin emission and that the main sources of error come from
$T_\mathrm{MB}\,(1,1;m)$ and $T_\mathrm{MB}\,(2,2;m)$. Defining $R \equiv T_\mathrm{MB}\,(2,2;m)/T_\mathrm{MB}\,(1,1;m)$, the
relative error is $\frac{\delta R}{R} = \sqrt{\big(\frac{\delta T_\mathrm{MB}\,(1,1;m)}{T_\mathrm{MB}\,(1,1;m)}\big)^2+
\big(\frac{\delta T_\mathrm{MB}\,(2,2;m)}{T_\mathrm{MB}\,(2,2;m)}\big)^2}$, with $\delta T_\mathrm{MB}\,(1,1;m)$ and $\delta
T_\mathrm{MB}\,(2,2;m)$ given directly by the hyperfine fit. Then, the error in the rotational temperature has been estimated as,


\begin{equation}
\delta T_\mathrm{rot}^{21}=
\frac{-41.5}{\ln^2(0.283R)}
\frac{\delta R}{R}.
\label{eeTrot}
\end{equation}

Then, as a test for the previous estimate of the error in $\Trot$, we estimated the opacity from the ratio of the main
line intensity to the inner satellites averaged intensity, $T_{\mathrm{MB}}\,(1,1,is)$, following Ho \& Townes (1983),

\begin{equation}
\frac{T_\mathrm{MB}\,(1,1;m)}{T_\mathrm{MB}\,(1,1;is)}=
\frac{1-e^{-\tau_\mathrm{m}}}{1-e^{-\tau_\mathrm{m}/3.6}},
\label{eeTrot}
\end{equation}
and derived $\Trot$ with this estimate of the opacity. The opacities derived from this method are systematically lower but
compatible with the values derived from the hyperfine fit, and the rotational temperature obtained agrees with the values derived from the
\nh\,(1,1) hyperfine method used in this work. We note that the opacity inferred from the hyperfine fit seems to be more reliable than the
opacity from the ratio of the main line to the satellites \emph{when} the width of the magnetic hyperfine components is comparable to the
their separation in velocity (\eg\ Anglada \et\ 1995), which is 0.11--0.53~\kms. This is probably the case of our region. However, since
the observations reported here were carried out with a spectral resolution in some cases lower than the intrinsic line width, the hyperfine
fits must be regarded with caution and for this reason we compared them with the ratio of the main line to the satellites method.





\begin{thebibliography}{}


\bibitem[Adams \& Myers(2001)]{adams2001}
Adams, F.~C., \& Myers, P.~C. \
2001, \apj, 553, 744

\bibitem[Andr\'e (1996)]{andre1996} 
Andr\'e, P. \
1996, in Taylor A.R., Paredes, J.M., eds, ASP Conf. Ser. Vol 93, Radio
Emission from the Stars and the Sun. Astrom. Soc. Pac., San Francisco, p. 273



\bibitem[Anglada et al.(1995)]{anglada1995} 
Anglada, G., Estalella, R., Mauersberger, J., Torrelles, J.~M., Rodr\'{\i}guez, L.~F.,
Cant\'o, J., Ho, P.~T.~P., \& D'Alessio, P. \
1995, \apj, 443, 682

\bibitem[Anglada et al.(1997)]{anglada1997}
Anglada, G., Sep\'ulveda, I., \& G\'omez, J.~F.\
1997, A\&AS, 121, 255


\bibitem[Anglada et al.(1998)]{anglada1998}
Anglada, G., Villuendas, E., Estalella, R., Beltr\'an, M.~T., Rodr\'{\i}guez, L.~F.,
Torrelles, J.~M. \& Curiel, S.\
1998, \apj, 116, 2953








\bibitem[Beltr\'an et al.(2001)]{beltran2001} 
Beltr\'an, M.~T., Estalella, R., Anglada, G., Rodr\'{\i}guez, L.~F. \&
Torrelles, J.~M.\ 2001, \apj, 121, 1556



\bibitem[Beltr\'an et al.(2006)]{beltran06}
Beltr\'an, M. T., Girart, J. M., \& Estalella, R.\
2006, \aap, 457, 865




\bibitem[Benson \& Myers(1983)]{benson1983} 
Benson, P.~J., \& Myers, P.~C.\
1983 \apj, 270, 589

\bibitem[Benson \& Myers(1989)]{benson1989} 
Benson, P.~J., \& Myers, P.~C.\
1989 \apjs, 71, 89


\bibitem[Briggs(1995)]{briggs1995}
Briggs, D.\
1995, PhD Thesis, New Mexico Inst. of Mining and Technology

\bibitem[Brooke et al.(2007)]{brooke2007} 
Brooke, T.~Y., Huard, T.~L., Bourke, T.~L., Boogert, A.~C.~A., Allen, L.~E., Blake, G.~A., Evans, N.~J.,~II, Harvey, P.~,M., Koerner, D.~W., Mundy, L.~G. \et\ \
2007, \apj, 655, 364

\bibitem[Carilli \& Holdaway(1997)]{carilli1997}
Carilli, C.~L., \& Holdaway, M.~A.\
1997, Millimeter Array Technical Memo 173 (NRAO)

\bibitem[Carrasco-Gonz\'alez et al.(2008)]{carrasco-gonzalez2008}
Carrasco-Gonz\'alez, C., Anglada, G., Rodr\'{\i}guez, L.~F., Torrelles, J.~M., Osorio, M., \& Girart, J.~M.\
2008, \apj, 676, 1073

\bibitem[Chandler \& Richer(2000)]{chandler2000}
Chandler, C.~J., \& Richer, J.~S.\
2000, \apj, 530, 851

\bibitem[Chen et al.(2008)]{chen2008a}
Chen, X., Bourke, T.~L., Launhardt, R., \& Henning, T\
2008, \apj, 686, L107

\bibitem[Chen et al.(2009)]{Chen2009}
Chen, X., Launhardt, R., \& Henning, T.\
2009, \apj, 691, 1729

\bibitem[Choi et al.(2008)]{choi08}
Choi, M., Hamaguchi, K., Lee, J-.E., \& Tatematsu, K.\
2008, \apj, 687, 406

\bibitem[Clarke et al.(2000)]{clarke2000} 
Clarke, C.~J., Bonnell, I.~A., \& Hillenbrand. L.~A.\
2000, in \textit{Protostars \& Planets IV},
eds. V. Mannings, A. P. Boss, \& S. S. Russell, (Tucson: University of Arizona Press), p. 151 

\bibitem[Condon et al.(1998)]{condon98}
Condon, J.~J., Cotton, W.~D., Greisen, E.~W., Yin, Q.~F., Perley, R.~A., Taylor, G.~B., \& Broderick, J.~J.\
1998 \aj, 115, 1693



\bibitem[de Gregorio-Monsalvo et al.(2005)]{deGregorio05} 
de Gregorio-Monsalvo, I., Chandler, C.~J., G\'omez, J.~F.,
Kuiper, T.~B.~H., Torrelles, J.~M., Anglada, G. \
2005, \apj, 628, 789

\bibitem[de Gregorio-Monsalvo et al.(2006)]{deGregorio06}
de Gregorio-Monsalvo, I., G\'omez, J.~F., Su\'arez, O., Kuiper,
T.~B.~H., Rodr\'{\i}guez, L.~F., Jim\'enez-Bail\'on, E.\
2006, \apj, 642, 319

\bibitem[Djupvik et al.(2006)]{djupvik2006}
Djupvik, A.~A., Andr\'e, Ph., Bontemps, S., Motte, F., Olofsson, G., G\aa lfalk, M., Flor\'en,
H.-G.\
2006, \aap, 458, 789 







\bibitem[Estalella et al.(2009)]{estalella2009}
Estalella, R., Palau, A., Girart, J.~M., Beltr\'an, M.~T., Osorio, M., Ho, P.~T.~P, \& Anglada, G.\
2009, \aap, submitted 


\bibitem[Felli et al.(1992)]{felli1992} 
Felli, M., Palagi, F., \& Tofani, G.\
1992, \aap, 255, 293 

\bibitem[Forbrich et al.(2009)]{forbrich2008}
Forbrich, J., Stanke, T., Klein, R., Henning, T., Menten, K.~M., Schreyer, K., Posselt, B.\
2009, \aap, 493, 547






\bibitem[Garay et al.(2003)]{garay03}
Garay G., Brooks K.~J., Mardones D., \& Norris R.~P.\
2003, \apj, 587, 739

\bibitem[Gibb(1999)]{gibb1999} 
Gibb, A.~G.\
1999, \mnras, 304, 1



\bibitem[Girart et al.(2000)]{girart00}
Girart, J.~M., Estalella, R., Ho, P.~T.~P., \& Rudolph, A. L. \
2000, \apj, 539, 763

\bibitem[Girart et al.(2009)]{girart2009}
Girart, J.~M., Rao, R., \& Estalella, R. \ 
2009, \apj, 694, 56



\bibitem[G\'omez et al.(1993)]{gomez1993} 
G\'omez, M., Hartmann, L., Kenyon, S.~J., \& Hewett, R.\
1993, \aj, 105, 1927 


\bibitem[Greve et al.(1998)]{greve98}
Greve, A., Kramer, C., \& Wild, W.\
1998, A\&AS, 133, 271

\bibitem[Gutermuth et al.(2008)]{gutermuth2008}
Gutermuth, R.~A., Bourke, T.~L., Allen, L.~E., Myers, P.~C., Megeath, S.~T., Matthews, B.~C. \et\
2008, \apj, 673, L151


\bibitem[Han et al.(1998)]{han1998} 
Han, F., Mao, R.~Q., Lu, J., Wu, Y.~F., Sun, J., Wang, J.~S., Pei, C.~C,
Fan, Y., Tang, G.~S. \& Ji, H.~R.\ 1998, \aaps, 127 181

\bibitem[Harju et al.(1993)]{harju1993}
Harju, J., Walmsley, C.~M., \& Wouterloot, J.~G.~A.\
1993, \aaps, 98, 51


\bibitem[Hartmann(1998)]{hartmann1998} 
Hartmann, L.\
1998, in \textit{Accretion Processes in Star Formation}, eds. A.
King, D. Lin, S. Maran, J. Pringle, \& M. Ward, Cambrigde: Cambridge University
Press, p.8

\bibitem[Hartquist, Williams, \& Viti(2001)]{Hartquist2001}
Hartquist, T.~W., Williams, D.~A., \& Viti, S.\
2001, \aap, 369, 605

\bibitem[Herbst \& Klemperer(1973)]{herbst1973} 
Herbst, E., Klemperer, W.\
1973, \apj, 185, 505


\bibitem[Ho \& Townes(1983)]{ho1983} 
Ho, P.~T.~P., \& Townes, C.~H.\
1983, \araa, 21, 239

\bibitem[Hogerheijde et al.(2000)]{Hogerheijde00}
Hogerheijde, M.~R. \& Sandell, G.\
2000, \apj, 534, 880

\bibitem[Hotzel et al.(2004)]{hotzel2004}
Hotzel, S., Harju, J., \& Walmsley, C.~M.\
2004, \aap, 415, 1065

\bibitem[Huard et al.(1999)]{huard1999}
Huard, T.~L., Sandell, G., \& Weintraub, D.~A.\
1999, \apj, 526, 833


\bibitem[Itoh et al.(1996)]{itoh1996}
Itoh, Y., Tamura, M., \& Gatley, I.\
1996, \apj, 465, L129



\bibitem[Kenyon et al.(1993)]{Kenyon93}
Kenyon, S.~J., Calvet, N., \& Hartmann, L.\
1993, \apj, 414, 676 


\bibitem[Kun(2008)]{kun2008}
Kun, M.\
2008, \textit{Star Forming Regions in Cassiopeia}
Handbook of Star Forming Regions, Volume I: The Northern Sky ASP Monograph Publications, Vol.~4.~Edited by Bo Reipurth, p.240





\bibitem[Lada(1999)]{lada1999} 
Lada, C.~J.\
1999, in \textit{The Origin of Stars and Planetary Systems}, eds. C. J. Lada, and N. D. Kylafis, Kluwer Acad.
Publ., p. 143 

\bibitem[Lada \& Lada(2003)]{lada2003} 
Lada, C.~J., \& Lada, E.~A.\
2003, \araa, 41, 57 

\bibitem[Lada et al.(1993)]{lada1993}
Lada, E.~A., Strom, K.~M., \& Myers, P.~C.\
1993, in \textit{Protostars \& Planets III}, eds. E. H. Levy and J. I. Lumine, Tucson: University of Arizona Press
p.245

\bibitem[Lee et al.(2006)]{lee2006} 
Lee, J-E., Di Francesco, J., Lai, S-P, Bourke, T.~L., Evans II, N. J., Spiesman, B., Myers, P.~C., Allen, L.~E., Brooke, T.~Y., Porras, A., \& Wahhaj, Z.\
2006, \apj, 648, 49 

\bibitem[Lehtinen et al.(2003)]{lehtinen2003}
Lehtinen, K., Harju, J., Kontinen, S., \& Higdon, J.~L.\
2003, \aap, 401, 1017

\bibitem[Matsuyanagi et al.(2006)]{matsuyanagi2006}
Matsuyanagi, I., Itoh, Y., Sugitani. K., \et\
2006, PASJ, 58, L29

\bibitem[Moriarty-Schieven et al.(2006)]{moriarty2006} 
Moriarty-Schieven, G.~H., Johnstone, D., Bally, J., \& Jenness, T.\
2006, \apj, 645,357

\bibitem[Motte \& Andr\'e(2001)]{motte2001}
Motte, F., \& Andr\'e, P. \
2001, \aap, 365, 440

\bibitem[Motte et al.(1998)]{motte98} 
Motte, F., Andr\'e, P., \& Neri, R.\
1998, \aap, 336, 150


\bibitem[Ohishi et al.(1992)]{ohishi92}
Ohishi, M., Irvine, W.~M., \& Kaifu, N.\
1992, in Astrochemistry of Cosmic Phenomena, ed. P.D. Singh (Dordrecht: Kluwer), 171

\bibitem[Ojha et al.(2004)]{ojha2004} 
Ojha, D.~K., Tamura, M., Nakajima, Y., \& Fukagawa M.\
2004, \apj, 608, 797


\bibitem[Ossenkopf \& Henning(1994)]{ossenkopf94}
Ossenkopf, V., \& Henning, Th.\
1994, \aap, 291, 943



\bibitem[Pauls et al.(1983)]{pauls1983}
Pauls, A., Wilson, T.~L., Bieging, J.~H., \& Martin, R.~N.\
1983, \aap, 124, 23

\bibitem[Pfalzner et al.(2008)]{pfalzner2008}
Pfalzner, S., Tackenberg, J., \& Steinhausen, M. \
2008, \aap, 487, L45


\bibitem[Rodr\'{\i}guez et al.(1999)]{rodriguez1999} 
Rodr\'{\i}guez, L.~F., Anglada, G. \& Curiel, S.\
1999, \apjs, 125, 427

\bibitem[Rodr\'{\i}guez et al.(2005)]{rodriguez05}
Rodr\'{\i}guez L.~F., Garay G., Brooks K.~J.,\&  Mardones D.\
2005, \apj, 626, 953

\bibitem[Rodr\'{\i}guez \& Reipurth(1989)]{rodriguez89}
Rodr\'{\i}guez L.~F., \& Reipurth, B.\
1989, Rev. Mex. Astron. Astrofis. 17, 59



\bibitem[Sep\'ulveda(1993)]{sepulveda1993} 
Sep\'ulveda, I.\
1993, Master Thesis. University of Barcelona


\bibitem[Sep\'ulveda(2001)]{sepulveda2001} 
Sep\'ulveda, I.\
2001, PhD Thesis. Universitat de Barcelona

\bibitem[Shimajiri et al.(2008)]{shimajiri2008}
Shimajiri, Y., Takahashi, S., Takakuwa, S., Saito, M., \& Kawabe, R.\
2008, \apj, 683, 255

\bibitem[Shu et al.(1987)]{shu1987} 
Shu, F.~H., Adams, F.~C., \& Lizano, S.\
1987, \araa, 25, 23

\bibitem[Sweitzer et al.(1978)]{Sweitzer1978}
Sweitzer, J. S.\
1978, \apj, 225, 116

\bibitem[Swift \& Welch(2008)]{swift2008}
Swift, J.~J., \& Welch, W.~J.\
2008, ApJS, 174, 202


\bibitem[Tafalla et al.(2004)]{tafalla2004}
Tafalla, M., Myers, P.~C, Caselli, P., \& Walmsley, C.~M.\
2004, \aap, 416, 191


\bibitem[Teixeira et al.(2007)]{teixeira2007}
Teixeira, P.~S., Zapata, L.~A., \& Lada, C.~J.\
2007, \apj, 667, L179




\bibitem[Ungerechts et al.(1986)]{ungerechts86}
Ungerechts, H., Winnewisser, G., \& Walmsley, C.~M.\
1986, \aap, 157, 207



\bibitem[Yang et al.(1990)]{yang1990} 
Yang, J., Fukui, Y., Umemoto, T. \& Ogawa, H.\
1990, \apj, 362, 538


\bibitem[Yokogawa et al.(2003)]{yokogawa03} 
Yokogawa, S., Kitamura, Y., Momose, M., \& Kawabe, R.\
2003, \apj, 595,266


\end{thebibliography}
\end{document}